\documentclass[10pt,journal]{IEEEtran}

\usepackage[utf8]{inputenc}
\usepackage[T1]{fontenc}

\usepackage{courier}
\usepackage{microtype}

\usepackage[noadjust]{cite}
\usepackage[hyphens]{url}

\usepackage{graphicx}

\usepackage[caption=false,font=footnotesize]{subfig}
\usepackage[inline]{enumitem}

\usepackage{textcomp}
\usepackage{siunitx}
\DeclareSIUnit{\nothing}{\relax}
\DeclareSIUnit\bps{bps}
\DeclareSIUnit\bits{bits}
\DeclareSIUnit\Byte{B}
\DeclareSIUnit\byte{byte}
\DeclareSIUnit\bytes{bytes}

\usepackage{balance}

\usepackage[capitalise,nameinlink]{cleveref}
\crefname{equation}{}{}
\Crefname{figure}{Fig.}{Figs.}
\crefname{line}{line}{lines}

\begin{document}

\bstctlcite{BSTcontrol}

\title{Computing Without Borders: The Way\\Towards Liquid Computing}

\author{Marco~Iorio,
        Fulvio~Risso,
        Alex~Palesandro,
        Leonardo~Camiciotti,
        and~Antonio~Manzalini
\thanks{M. Iorio and F. Risso are with Politecnico di Torino, Torino, Italy. 
E-mail: \{name\}.\{surname\}@polito.it. A. Palesandro was with Politecnico di Torino at the time of this work. L. Camiciotti is with Consorzio TOP-IX, Torino, Italy. A. Manzalini is with TIM, Torino, Italy.}
\thanks{This work has been published in IEEE Transactions on Cloud Computing: \protect\url{https://doi.org/10.1109/TCC.2022.3229163}.}
\thanks{\copyright 2022 IEEE. Personal use of this material is permitted.  Permission from IEEE must be obtained for all other uses, in any current or future media, including reprinting/republishing this material for advertising or promotional purposes, creating new collective works, for resale or redistribution to servers or lists, or reuse of any copyrighted component of this work in other works.}
}

\maketitle

\begin{abstract}
Despite the de-facto technological uniformity fostered by the cloud and edge computing paradigms, resource fragmentation across isolated clusters hinders the dynamism in application placement, leading to suboptimal performance and operational complexity.
Building upon and extending these paradigms, we propose a novel approach envisioning a transparent continuum of resources and services on top of the underlying fragmented infrastructure, called \textit{liquid computing}.
Fully decentralized, multi-ownership-oriented and intent-driven, it enables an overarching abstraction for improved applications execution, while at the same time opening up for new scenarios, including resource sharing and brokering.
Following the above vision, we present \emph{liqo}, an open-source project that materializes this approach through the creation of dynamic and seamless Kubernetes multi-cluster topologies.
Extensive experimental evaluations have shown its effectiveness in different contexts, both in terms of Kubernetes overhead and compared to other open-source alternatives.
\end{abstract}


\section{Introduction}\label{sec:introduction}

\IEEEPARstart{I}{n} the last years, containerization has increasingly gained popularity as a lightweight solution to package applications in an interoperable format~\cite{PAHL2015}, independently of the target infrastructure.
This uniform substratum paved the way for the cloud native revolution, with novel applications shifting their focus from single servers to entire data centers, and where dedicated orchestrators manage the lifecycle of microservice applications.
As of today, Kubernetes emerged as the de-facto open-source framework for container orchestration, bridging the semantic gaps across competing infrastructure providers~\cite{CNCF2021}.
With the rise of the edge and fog computing paradigms~\cite{GARCIALOPEZ2015,SHI2016,BONOMI2012} as solutions accounting for geographical closeness, reduced latency and improved privacy, the same approaches are being progressively extended towards smaller data centers at the network border, benefiting from uniform primitives to foster service agility.

Despite the emergence of common interfaces for applications orchestration being key towards a real \emph{edge to cloud continuum}~\cite{MILOJICIC2020,BARESI2019}, industry-standard approaches handle each infrastructure as a multitude of (connected) isolated silos instead of a unique virtual space. This leads to a sub-optimal fragmented view of the overall available resources, preventing the seamless deployment of fully distributed applications.
Indeed, edge data centers cannot depend on a single centralized control plane, for resiliency (i.e., preventing failure propagation in case of network partitioning) and performance reasons, as orchestration platforms typically suffer if nodes are geographically spread over high-latency WANs~\cite{MULUGETA2020,LARSSON2020A,OSMANI2021}.
Besides the edge landscape, resource fragmentation affects also larger data centers, with many companies increasingly witnessing the cluster sprawl phenomenon~\cite{GARTNER2020,D2IQ2021}.
This trend finds its roots in scalability concerns, in the hybrid-cloud (i.e., the combination of on-premise and public cloud) and multi-cloud approaches~\cite{NIST2011}, which aim for high availability, geographical distribution and cost-effectiveness, while granting access to the breadth of capabilities offered by competing cloud providers.
Additionally, non-technical requirements such as law regulations, mergers and acquisitions, physical isolation policies and separation of concerns contribute to the proliferation of clusters.

Fragmentation also hinders the potential dynamism in the workload placement~\cite{YOUSAFZAI2017,RAJKUMAR2018,MAENHAUT2019}, forcing each application to be assigned upfront to a specific infrastructure.
No resource compensation is ever possible, hence preventing jobs from transparently moving from an overloaded cluster, e.g., due to unexpected spikes of requests, to another one, underused and potentially offering better performance.
At the same time, the deployment of complex applications composed of multiple microservices, each one with specific requirements (e.g., low latency, high computational power, access to specialized hardware, \dots), as well as the enforcement of proper geographical distribution and high-availability policies, requires the interaction with different infrastructures.
However, this prevents to rely on the single point of control abstraction, which would allow to coordinate the deployment of arbitrarily complex applications across the entire resource continuum, no matter how many nodes and clusters it is composed of.

Accounting for these demands, in this paper we advocate the opportunity for a novel architectural paradigm: \emph{liquid computing}\footnote{This term was first coined in 2014 by InfoWorld~\cite{INFOWORLD2014} as a synonym of pervasive computing, i.e., the capability of keep working on a given task across multiple devices such as PCs and tablets. This paper refers to a broader concept, which encompasses the creation of a resource continuum composed of cloud and edge infrastructures, on-premise clusters, as well as, in its widest form, single end-user and IoT devices.}, which builds upon and extends the well-established cloud and edge computing approaches towards an endless \textit{computing continuum}.
Then, we present a first real implementation of a software framework enabling a continuum of computational resources and ready-to-consume services spanning across multiple physical infrastructures.
Overall, the resulting computing domain abstracts away the specificity of each cluster, presenting to the final users, either actively participating as actors or simply renting off-the-shelf services, a \emph{unique} and \emph{borderless} pool of available resources, the so-called \emph{big cluster}.
Thanks to this abstraction, applications are no longer constrained in a specific silo, but free to \emph{fly} in the entire infrastructure, selecting the most appropriate location depending on its requirements (e.g., a user facing service may be replicated at the edge to account for low latency, while another might be constrained to European infrastructures to comply with GDPR), and the available resources, while retaining full compatibility (hence, models, tools, and commands) with vanilla Kubernetes. 

The remainder of the paper is organized as follows. \Cref{sec:vision} discusses our liquid computing vision, with its key pillars detailed in \Cref{sec:pillars}. \Cref{sec:liqo} describes the main characteristics of \emph{liqo}\footnote{https://liqo.io}, an open-source project which fosters this idea by enabling dynamic and seamless Kubernetes multi-cluster topologies, with the most relevant implementation aspects detailed in \Cref{sec:implementation}. \Cref{sec:validation} presents its experimental evaluation, while \cref{sec:related-work} discusses related approaches. Finally, \Cref{sec:conclusions} draws the main conclusions.
\section{The Liquid Computing Vision}\label{sec:vision}

We envision liquid computing as a \emph{continuum of resources and services} allowing the seamless and efficient deployment of applications, independently of the underlying infrastructure.
We present here the main characteristics of liquid computing, followed by the most significant deployment scenarios.

\subsection{Main characteristics} \label{sec:vision:characteristics}

We believe this paradigm shall be composed of the following four distinguishing characteristics.

\subsubsection{Intent-driven}\emph{A consumer can assign to each workload the desired execution constraints through high-level policies, without knowing about the infrastructural details.}
Overall, liquid computing brings the \emph{cattle service model}~\cite{BIAS2012} to a greater scale.
Similarly to servers in a data center, with no one caring about where each task is executed, as long as requirements are fulfilled, this paradigm blurs the cluster borders so that users are relieved from selecting a specific infrastructure for their applications.
Yet, different clusters are definitely associated with different properties (e.g., in terms of geographical location and security characteristics) and, indeed, this is one of the main driving reasons behind cluster sprawling.
Thus, it is of utmost importance the adoption of an \emph{intent-driven} approach, allowing final users to enrich each workload with a set of high-level policies to express the associated constraints (e.g., geographical locality and spreading, costs, capabilities, \dots); automated schedulers shall select the best execution place across the entire border-less infrastructure, depending on the available resources and enforcing in concert the user-specified policies. 
Yet, we deem at the same time the resource continuum abstraction to enable more contextualized scheduling decisions (given the knowledge about the entire infrastructure), allowing for further optimizations and better scalability compared to the siloed approach.

\subsubsection{Decentralized architecture}\emph{The resource continuum stems from a peer-to-peer approach, with no central point of control and management entities, as well as no intrinsically privileged members.}
Following a decentralized and peer-based model like the Internet, the liquid computing approach fosters the coexistence of multiple actors, including larger cloud providers, smaller, territory-linked enterprises and possibly even small office/home owners.
Indeed, each entity can autonomously and dynamically decide who to peer with (hence, share resources), similarly to the concept of \emph{Autonomous Systems} in the Internet inter-domain routing.
A dynamic \emph{discovery and peering} protocol is in charge of the automatic identification of available peers and the negotiation of peering contracts based on the demands and offers of each actor, along with their specific constraints; optionally, the above operations could also be delegated to an intermediate dedicated entity such as a \emph{broker}.
No sensitive information disclosure is mandated (e.g., infrastructural setup), with the entire process possibly involving only the request for a certain amount of abstract resources (e.g., CPU, memory, \dots) and the offer of available ones, together with the associated cost. 
Besides peering establishment, decentralization also concerns the preserved ability of each cluster to evolve independently, thanks to the local orchestration logic, and the support for the creation of arbitrary topologies, with different points of entry for the deployment of different workloads.

\begin{figure*}
    \centering
	\subfloat[]{\includegraphics{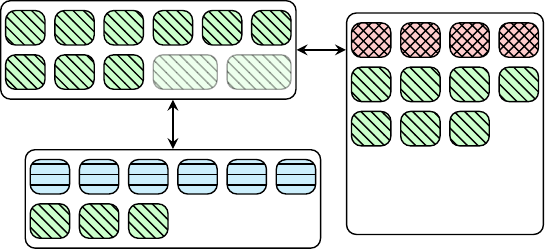}\label{fig:deployment-scenarios-resource-sharing}}
	\hfill
	\subfloat[]{\includegraphics{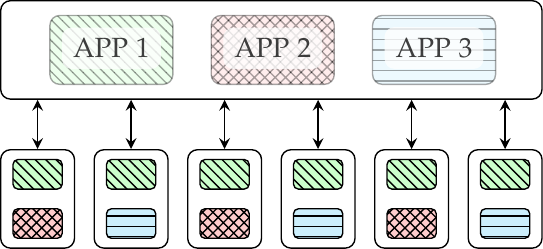}\label{fig:deployment-scenarios-super-cluster}}
	\hfill
	\subfloat[]{\includegraphics{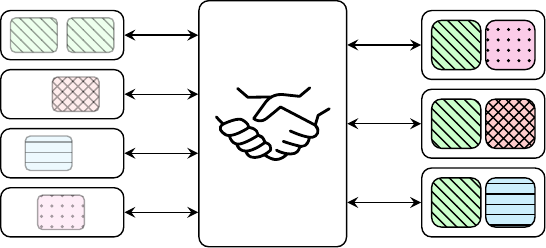}\label{fig:deployment-scenarios-resource-exchange-point}}
	\caption{A graphical representation of the three deployment scenarios fostered by liquid computing:	\protect\subref{fig:deployment-scenarios-resource-sharing} elastic cluster, allowing applications to spill over to federated infrastructures,
    \protect\subref{fig:deployment-scenarios-super-cluster} super cluster, providing an abstraction to control multiple independent infrastructures, and
	\protect\subref{fig:deployment-scenarios-resource-exchange-point} brokering cluster, enabling third parties to match resource demands and offers.}
	\label{fig:deployment-scenarios}
\end{figure*}

\subsubsection{Multi-ownership}\emph{Each actor maintains the full control of his own infrastructure, while deciding at any time how many resources and services to share and with whom.}
Although single clusters are expected to be under the control of a single entity, the entire \emph{resource ocean} would likely span across different administrative domains. 
Once a new peering is established, the control plane of the target infrastructure is in charge of configuring the appropriate isolation primitives (e.g., resource quota, network and security policies, \dots), based on the underlying orchestration capabilities, to enforce the shared resource slice and prevent noisy neighbors phenomena.
Specifically, we foresee a \emph{shared security responsibility model}, with the provider responsible for the creation of well-defined sandboxes and the possible provisioning of additional security mechanisms (e.g., secure storage) negotiated at peering time.
Requesters, on the other hand, are expected to take measures to fortify their applications (similarly to public cloud computing) and to configure for each sensitive component the appropriate policies to ensure it is scheduled on security compliant infrastructures only (e.g., private data is processed locally).

\subsubsection{Fluid topology}\emph{Members can join and leave at any time, while spanning across the entire range of infrastructure sizes, from enterprise-grade data centers to IoT and personal devices.}
Generalizing traditional federation approaches, liquid computing aims at supporting highly dynamic scenarios, with frequent and unexpected (or, in other scenarios, explicitly desired) connections and disconnections. 
Besides spanning across public and private data centers, as well as edge clusters, the resource continuum possibly encompasses also single devices.
This would include IoT, industrial and domestic scenarios, all characterized by a multitude of independent appliances typically dedicated to specific tasks (e.g., machine tools control, home automation, monitoring, \dots), which could greatly benefit from this paradigm, transparently leveraging the shared resources to offload computations and extend their capabilities~\cite{RAJKUMAR2018}.

\subsection{Deployment scenarios} \label{sec:vision:scenarios}
Overall, we deem three deployment scenarios to be mostly enabled and fostered by liquid computing (\cref{fig:deployment-scenarios}).

\subsubsection{Elastic cluster}
The liquid computing paradigm reduces the fragmentation of scattered clusters thanks to the possibility of transparently leveraging the resources available in other locations, which enables to balance and absorb load spikes (\textit{cloud bursting}).
This definitely applies to edge-computing scenarios, whose pervasiveness is typically achieved at the expense of the computational capacity, but is visible also in more traditional cloud contexts in which multiple clusters are active. For example, elastic clusters can be used to support multi-cloud strategies (no single vendor lock-in) and geographically distributed multi-cluster deployments (e.g., due to cost optimization, latency, redundancy or legislative concerns).
Resource sharing can involve both on-premise and public data centers, hence deploying latency-sensitive applications close to the final users, while benefiting from the virtually infinite computational capacity featured by larger infrastructures for resource intensive tasks.
Thanks to the decentralized approach and the support for peering contracts, resource sharing can occur also when clusters are under the control of different organizations, i.e., they belong to different administrative domains.

\subsubsection{Super cluster} 
We are currently witnessing scenarios in which a company owns hundreds, or even thousands, of small clusters, often located at the edge of the network, such as a large telecom operator.
In this context, the liquid computing paradigm enables a higher-level, \emph{super cluster} abstraction, representing the single point of entry that can transparently deploy and control applications hosted by the entire infrastructure.
Although apparently centralized, each level-2 cluster actually maintains its own local orchestration logic, hence being resilient to network outages and preventing possible synchronization issues arising with relatively high latency WAN links~\cite{MULUGETA2020,LARSSON2020A,OSMANI2021}.
In addition, thanks to the super cluster, the administrative burden is greatly reduced, enabling a borderless orchestration that geographically distributes and controls the tasks from a single point of entry, without the necessity of an explicit interaction with the underlying clusters.
This scenario facilitates the automatic migration of applications from one cluster to another, which helps when dealing with disaster recovery, infrastructure interventions, scaling, or placement optimization.
In addition, it greatly simplifies the replication of jobs across clusters (hence management, monitoring and troubleshooting) as it leverages existing primitives that operate on nodes of the super cluster; for example, an operator can easily replicate the same service on a subset of its edge clusters in order to serve all the end users present in their close vicinity.
Finally, this approach can be combined with the elastic cluster for increased dynamism, thus benefiting from the single point of entry for workloads replication, while enabling at the same time the offloading of bursty workloads from the edge to the cloud.

\subsubsection{Brokering cluster} Complex applications require complex infrastructural setups, accounting for both resiliency and performance. While larger cloud providers could theoretically offer a sufficiently wide catalog of services to satisfy most demands, relying on a single vendor increases lock-in and potentially leads to cost inefficiencies.
In this context, we believe that liquid computing could foster the creation of new \emph{Resource and Service Exchange Points} (RXPs), with third party entities behaving as brokers between consumers and providers.
Consumers would then only need to peer (both technologically and contractually) with a single RXP to immediately benefit from the entire set of resources (cloud and edge data centers, \dots) and ready-to-use services therein offered.
This would reduce the complexity and the operational costs especially for smaller companies, lacking the bargaining power of larger enterprises.
Resource providers, at the same time, would be encouraged in participating, to easily reach a wider turnout of interested customers.
Additionally, even small actors, such as the ones operating at the edge of the network, would be enabled to participate in the edge-cloud market, possibly in a fair competition with far larger giants that may not have enough resources to serve a given geographical area, in a way that looks similar to the energy market in which millions of tiny producers are aggregated by larger buyers.
\section{The Liquid Computing Pillars}\label{sec:pillars}

This section presents the main technical building blocks required to materialize the liquid computing vision, assuming Kubernetes as the orchestration platform leveraged by the underlying clusters. 
In fact, in our opinion Kubernetes represents a key enabler for liquid computing, thanks to its capillary diffusion in data centers of any size~\cite{CNCF2020}, as well as the support for single devices and IoT computing by means of lightweight distributions, such as \emph{k3s}~\cite{K3S}.
To this end, the recent Microsoft's backed \emph{Akri} project~\cite{GOLDENRING2020} goes even further, introducing an abstraction layer to dynamically interconnect to this platform the variety of sensors, controllers and MCU class devices typically present at the very edge of the network.
At the same time, Kubernetes can be easily extended through both custom APIs and logic, allowing to transparently integrating liquid-computing related aspects, as well as to semantically enrich the ecosystem and introduce new services that may be shared with peered clusters.
Hence, being the underlying platform (conceptually similar to an overarching operating system) the resource continuum is built upon.
Still, the key concepts are definitely more general, and can be applied with no particular difference to other orchestrators, such as OpenStack, or even to a mix thereof.

\subsection{Dynamic Discovery and Peering}

The first key enabler is the \emph{discovery and peering} function. 
It fosters the decentralized governance approach typical of a peer-to-peer architecture, preventing the need for central management entities and full administrative control over the entire infrastructure.
Additionally, it is responsible for the liquid computing dynamism, allowing for new peering relationships to be established and revoked at any time, compared to the manual coordination required by static federation approaches.
In this context, we define \emph{peering} a unidirectional resource and service consumption relationship, with one party (i.e., the consumer) granted the capability to offload tasks and/or consume services in a remote cluster (i.e., the provider), but not vice versa.
This allows for maximum flexibility in asymmetric setups, while transparently supporting bidirectional peerings through their combination.

Overall, this module deals with four main tasks.
\begin{enumerate*}[label=(\textit{\roman*)}]
\item \emph{Discovery}, to identify candidate clusters to peer with, including large enterprise domains, as well as possibly local independent appliances (e.g., IoT devices).
\item \emph{Authentication}: given the list of feasible candidates obtained during the previous step, optionally filtered through user-configured criteria, it is responsible for the establishment of a secure communication channel with each selected counterpart.
Still, resource offloading is not yet possible at this point, being the granted authorizations related to peering establishment steps only.
\item \emph{Resource negotiation}, involving the exchange of request and offer messages to identify the shortlist of clusters selected for resource offloading.
The entire process is policy-driven, with decision modules local to each cluster determining at each step whether to proceed with the negotiation or to abort the process.
As a representative example, an offering cluster might implement complex business logic to determine the appropriate prices based on current demands and available resources, accounting for resource brokering and reselling scenarios.
Consumers, on the other hand, may filter and rank the received offers by means of appropriate criteria, possibly including compliance with the request constraints, cost, additional attributes, past experience, and more.
The negotiation process culminates with the mutual agreement between a consumer and a provider. 
To this end, we envisage the adoption of smart contracts~\cite{ZHENG2020} to formalize the exchange in terms of money and resources, especially in case of inter-administrative domain peerings.
\item \emph{Peering finalization}: once resource negotiation is completed, the peering relationship needs to be finalized, leading to the exchange of the preparatory parameters required for subsequent computation offloading (e.g., network configurations, as analyzed in \cref{sec:pillars:network-continuum}), as well as the setup of isolation mechanisms and the granting of the suitable permissions in the target cluster.
\end{enumerate*}

\subsection{Hierarchical Resource Continuum}\label{sec:pillars:resource-continuum}

Once peering relationships have been established towards one or more targets, the local cluster gains logical access to remote resource slices.
Yet, these need to be properly exposed for application offloading through a continuum abstraction, while respecting the limited knowledge propagation and the multi-ownership constraints.
Moreover, we deem API transparency to be of utmost importance to foster its widespread adoption, thanks to the introduction of no disruption in well-established deployment and administration practices, as well as the immediate support for existing management solutions.

Being traditional clusters composed of multiple nodes, each one mapping to a physical server, we propose to represent peered clusters through \emph{local, virtual, big nodes}.
\emph{Local}, as attached to the consuming cluster; \emph{virtual}, since they abstract a set of remote resources possibly unrelated from the underlying hardware; and \emph{big}, being potentially much larger than classical nodes (in terms of available capabilities), as backed by an entire data center slice.
The node concept perfectly complies with the requirement of sharing limited information, hence abstracting peered clusters only in terms of the aggregated resources currently being shared, with no additional details regarding its actual internal configuration.
At the same time, it leads to overall better scalability, given the reduced amount of data synced among different clusters.
This approach opens up for two possible cluster models.
First, \emph{extended clusters}, encompassing a combination of traditional physical nodes (i.e., workers), and virtual ones. This could be suitable for the resource optimization and RXP consumer use-cases, to allow borrowing external computational capacity to overcome local limitations.
Second, \emph{virtual clusters}, characterized by the absence of local workers. Combining only virtual nodes, they provide a single point of control abstraction to simplify the deployment of applications on user-defined slices of the underlying infrastructure. 
Moreover, they represent a key enabler for resource brokering, aggregating the resources offered by multiple providers (each one mapped to a virtual node) for reselling.

The virtual node abstraction leads the underlying orchestration platform (e.g., Kubernetes) to consider the above nodes as valid scheduling targets, hence allowing traditional workloads to be transparently assigned to remote clusters.
No differences are perceived by the final users, who simply benefit from the enlarged amount of available resources.
This approach brings to a hierarchical representation of the resource continuum.
When a new workload is deployed in the local cluster, the scheduler first selects the optimal node for its execution. 
Then, if the target is a virtual node, the workload is remapped to the corresponding remote cluster, where it incurs in a second scheduling round to identify the physical server where it will be executed upon.
While considering a two-layer scheduling in this example, the approach can easily generalize to multiple levels if needed, depending on the number of virtual node redirections.
Hence, allowing scheduling decisions to occur at different abstraction layers, reducing the overall number of feasible candidates to consider at each step and potentially increasing the resulting accuracy.
Once more, compliance with standard Kubernetes APIs enables vanilla schedulers to deal out-of-the-box with extended clusters. 
However, custom scheduling logic might be appropriate in certain scenarios, allowing for further optimizations thanks to the knowledge about the semantics of the peering relationship (e.g., monetary costs, network characteristics, geographical distance, QoS).
In both cases, end-users can easily enforce domain-specific constraints through Kubernetes standard high-level policies (i.e., selectors and affinities) to assign workloads to slices of nodes and ensure replicas spreading.
Hence, sticking to an intent-driven approach, while requiring no modifications in standard application deployment workflows.

\subsection{Resource and Service Reflection}\label{sec:pillars:resource-reflection}

Each virtual node is responsible for its allocated workloads, whose execution is actually delegated to the remote cluster. Hence, selected control plane information should be present both in the local cluster (required to fulfill the requirement of the virtual node abstraction) and in the remote cluster (enabling the remote control plane to carry out its operations).
This introduces the \emph{resource reflection} concept: objects exist both in their \emph{native} form (i.e., in the local cluster), and in their \emph{shadow} form, remotely.
Indeed, applications most likely require accessory artifacts for proper execution (e.g., configurations, authorization tokens, network endpoints, etc.), which then need to be reflected in the target cluster.
The resource reflection logic enforces the transparent realignment between the two digital twins of the same artifact across the different domains, while ensuring the desired information opacity properties (i.e., omitting or masquerading data that should not be propagated) and resolving possible conflicts which may arise in the remote infrastructure (e.g., naming collisions, different underlying technologies, \dots).   
Overall, it shall support the propagation of both local modifications (e.g., the change in a user configuration) --- \emph{outgoing reflection} --- and of remote status changes (e.g., an application is being restarted due to a crash), hence allowing for proper inspection --- \emph{incoming reflection}.
Service endpoints represent one of the most important reflected information, enabling an application running on one cluster to be reachable (hence, consumable) from another cluster. This may require the close coordination of the network fabric (\cref{sec:pillars:network-continuum})  to disambiguate and transparently translate possible overlapped network addresses used in the communication flows.

The clever reflection of the required information is the key to achieve objectives such as \textit{robustness}, enabling clusters to evolve also in case of network disconnections, and \textit{scalability}, reducing the amount of synced data.

\subsection{Network Continuum}\label{sec:pillars:network-continuum}

According to the virtual nodes approach, different components of the same application may be spread across multiple clusters.
Still, the resource continuum, alone, is not sufficient to ensure their correct execution, as the various microservices most likely need to interact among each other. 

Orchestration platforms typically implement internal communication by means of private IP addresses, resorting to public ones only for user-facing services.
Hence, they are unsuitable for direct (pod-to-pod) cross-cluster interactions and require the introduction of an appropriate \emph{network fabric} responsible for the transparent communication between microservices, no matter where they are executed.
Accounting for the decentralized and dynamic approach fostered by liquid computing, with peers possibly joining and leaving at any time, the network fabric cannot rely on ahead-of-time knowledge for its establishment.
Indeed, it shall only require the cooperation between the two involved clusters, which negotiate the configuration parameters necessary to set up
\begin{enumerate*}[label={\emph{(\roman*)}}]
\item the secured communication channel and
\item the proper mechanisms to guarantee the any-to-any communication across the entire virtual cluster.
\end{enumerate*}
Being the interconnecting clusters potentially under the control of different administrative domains, it is likely conflicts may arise, e.g., in terms of overlapping IP addresses or underlying networking solutions.
The network fabric is expected to transparently handle all these issues, while virtually extending the local cluster network to the entire resource continuum, presenting a unique border-less addressing space.

Supposing a central cluster $C$ peered with $n$ others, we foresee two main network topologies for data plane communications. 
First, a hub and spoke topology, with $n$ direct interconnections between $C$ and all the leaves.
Conceptually simple, this solution requires all traffic between applications residing on peripheral clusters to flow through the central hub, potentially resulting in communication inefficiencies. 
Still, it may be appropriate when applications do not span across multiple remote clusters (e.g., the same application is replicated in multiple edge clusters), in case either the communication pattern or the underlying network match the star topology, as well as when traffic policies should be enforced from a single point of control. 
Second, an \emph{opportunistic mesh} topology, providing full connectivity between all clusters hosting applications potentially communicating between one another, to avoid traffic tromboning.

It is worth noting that peripheral clusters may in turn play the role of central clusters for different peering sessions, hence leading to completely dynamic and independent topologies, and potentially overlapped virtual clusters.

\begin{figure*}
    \centering
    \includegraphics{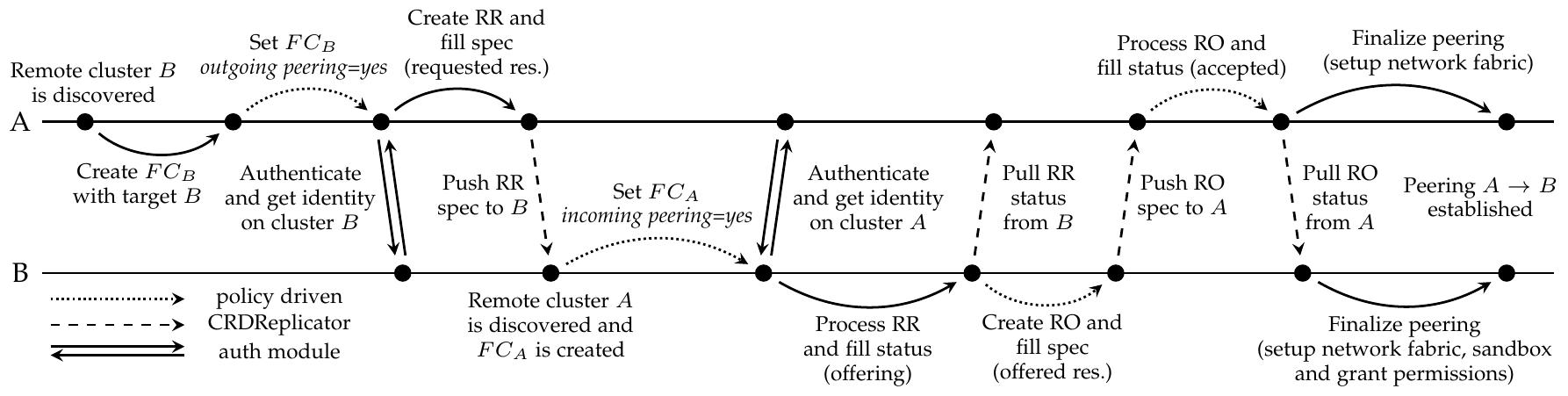}
    \caption{Schematic representation of the \emph{liqo} discovery and peering procedure. FC: ForeignCluster, RR: ResourceRequest, RO: ResourceOffer.}
    \label{fig:peering-process}
\end{figure*}

\subsection{Storage and Data Continuum}\label{sec:pillars:storage-continuum}
When an application is spread across multiple clusters, stateful workloads require the access to persistent storage locations, which implies a \textit{data continuum} across all the virtual cluster.
To this end, we foster the \emph{data gravity} approach borrowed from well-established practice in the Big Data world~\cite{FRITSCH2014}.
According to it, data attracts the associated workloads (i.e., introducing additional placement constraints) rather than vice versa, to ensure the best performance in terms of reduced network traffic and latency, as well as to enforce storage locality, which represents a possible strong requirement to comply with law regulations.
This paradigm allows also for the extension of traditional in-cluster stateful workloads replication mechanisms (e.g., databases) across the entire resource continuum, transparently achieving increased disaster recovery support.  
Information replication and synchronization might be further supported if more dynamism is desired, although requiring the exchange of data between potentially distant storage pools; hence, this is mostly suitable only in case of limited amounts of data.
\section{The Liqo Architecture}\label{sec:liqo}

This section presents \emph{liqo}, an open-source project\footnote{Source code is available at \url{https://github.com/liqotech/liqo}} fostering the liquid computing vision presented above. 
Acknowledging its wide diffusion and flexibility, \emph{liqo} builds upon and extends Kubernetes to enable dynamic and seamless multi-cluster topologies independently of the underlying infrastructural borders.
Overall, \emph{liqo} aims to introduce no modifications in standard Kubernetes APIs for application deployment and well-established management workflows, as well as to support a wide range of common infrastructures, with no constraints in terms of cluster type (i.e., on-premise or hosted by a cloud provider) and networking configurations (i.e., CNIs and IP addresses).
In the following, we detail its main architectural characteristics, while building a parallelism with the technical pillars presented in \cref{sec:pillars}.

\subsection{Discovering and Peering with Remote Clusters}
The \emph{liqo} discovery logic is responsible for the identification of possible remote clusters to peer with.
Accounting for different scenarios, \emph{liqo} supports (\textit{i}) manual configuration, and standard DNS-based Service Discovery~\cite{RFC6763}, leveraging both (\textit{ii}) conventional Unicast DNS, suitable for large enterprise domains, and (\textit{iii}) Multicast DNS, allowing dynamic on-LAN clustering of independent devices.
In all cases, the output is a remote network endpoint that can be later leveraged to start the authentication procedure: this information, along with the desired peering state (i.e., whether it should be established) and possible additional attributes is represented through a \emph{ForeignCluster} Custom Resource (CR).

The \emph{liqo} peering procedure (\cref{fig:peering-process}) starts once a given discovered cluster $B$ is selected as a desired target (i.e., \emph{outgoing peering} flag set in its ForeignCluster CR), either manually or through policies.
This procedure is entirely based on a Kubernetes-native logic, which consists in setting the proper resources in Kubernetes, possibly reflected in the other cluster by \emph{liqo}; however, a more traditional protocol-based approach could also be envisioned.
The first step involves the authentication module: the originating cluster $A$ generates a new private key locally, then sends a \emph{Certificate Signing Request (CSR)} and a pre-shared token to the remote endpoint. 
If authentication is granted, the remote module in $B$ proceeds signing the request, assigning $A$ just the bare permissions necessary during the peering establishment procedure, and eventually returning the generated certificate.
This approach completely integrates with standard Kubernetes permissions management, and it does not require any common certification authority among peering clusters.
Alternative solutions might be adopted in different scenarios, such as to comply with public cloud requirements.
In the end, the requesting cluster $A$ shall obtain a valid identity, later used to interact with the remote peer $B$.

The resource negotiation can now start. 
First, $A$ creates a new \emph{ResourceRequest} CR locally, to make explicit the desire to request computational resources and/or services to a remote cluster, and configures the content of its \emph{Spec} stanza to convey the desired information.
Then, the \emph{CRDReplicator}, which is responsible for the interaction among clusters through the replication of custom resources during the peering establishment, takes action, and it duplicates the CR on the remote cluster.
Once the ResourceRequest is received, $B$ automatically discovers cluster $A$, creates the corresponding \emph{ForeignCluster} representation and, in case it is willing to proceed with the resource negotiation (i.e., the \emph{incoming peering} field is set), it performs the symmetrical authentication procedure.
At the same time, the \emph{ResourceRequest} is processed by a custom logic and the outcome, along with possible additional parameters, are back-propagated through an update of its \emph{Status} stanza (eventually pulled by cluster $A$), which can be used to decline the peering request.
In case of acceptance, $B$ would emit a proper \emph{ResourceOffer} CR to convey the willingness of sharing a given amount of available resources/services, possibly at a given price, and replicate it to the requesting cluster $A$ through the CRDReplicator.
\emph{liqo} features resource negotiation based on a customizable amount of available resources, with the support for pluggable decision modules (e.g., for brokering scenarios).

Once a ResourceOffer is accepted, the new peering relationship can be finalized, establishing the inter-cluster network fabric (cf. \cref{sec:liqo:network-fabric}).
Additionally, cluster $B$ grants increased permissions to $A$, allowing for computation offloading in the target cluster, while configuring at the same time the appropriate isolation mechanisms in terms of network communication, security, resource usage, etc.
The established peering relationship is unidirectional, with $A$ being granted the possibility to leverage the resources offered by $B$, but not vice versa. 
Still, the reverse procedure can be later started by $B$, achieving a bidirectional peering.

\subsection{The Virtual Node Abstraction}\label{sec:liqo:virtual-node}

\emph{liqo} leverages the \emph{virtual node} concept to masquerade the resources shared by each remote cluster.
This solution allows the transparent extension of the local cluster, with the new capabilities seamlessly taken into account by the vanilla Kubernetes scheduler when selecting the best place for the workloads execution.
The virtual node abstraction is implemented through an extended version of the \emph{Virtual Kubelet} project~\cite{VIRTUALKUBELET}.
In Kubernetes, the \emph{kubelet} is the primary \emph{node agent}, responsible for registering the node with the control plane and handling the lifecycle of the \emph{pods} (i.e., the minimum scheduling unit, composed of one or many containers sharing the same network namespace) assigned to that node.
The virtual kubelet (VK) replaces a traditional kubelet when the controlled entity is not a physical node, allowing to control arbitrary objects through standard Kubernetes APIs.
Hence, it enables custom logic to handle the lifecycle of both the node itself and the pods therein hosted.

\subsubsection{Node lifecycle handling}

The first task handled by the VK regards the creation and the management of the \emph{virtual node} abstracting the resources shared by the remote cluster.
In particular, it aligns the node status (i.e., whether it is ready, as well as its size in terms of available resources) with respect to the negotiated configuration (i.e., ResourceOffer).
Periodic healthiness checks are performed to assess the reachability of the remote cluster, marking the virtual node as \emph{not ready} in case of repeated failures.
Upon this event, if disconnections are explicitly foreseen and shall be tolerated (e.g., to account for edge devices in harsh environments), existing workloads are allowed to evolve independently through the remote orchestration logic, with the virtual node no longer considered a valid scheduling target only for new applications.
Differently, in other scenarios, user configurations might require standard Kubernetes logic to proceed evicting all pods hosted on the failing cluster and reschedule them in a different location to ensure service continuity.

\subsubsection{Pod lifecycle handling}

Differently from a traditional kubelet, which starts the actual containers on the designated node, the \emph{liqo} VK implementation is conceptually responsible for mapping each operation to a corresponding \emph{twin} pod object in the remote cluster for actual execution, while possibly going through additional indirection levels in brokering scenarios. 
Still, remote status changes are automatically propagated to the respective local pods, hence allowing for proper monitoring and administrative inspection.
Advanced operations, including metrics and logs retrieval, as well as interactive command execution inside remote containers are transparently supported, to comply with standard troubleshooting operations.

\begin{figure}
    \centering
    \includegraphics{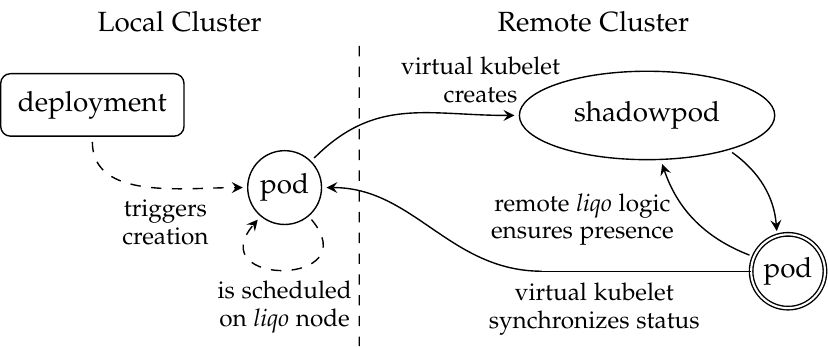}
    \caption{Schematic representation of the pod offloading workflow. Solid lines refer to \emph{liqo}-related tasks, while dashed ones to standard Kubernetes logic. Double circles indicate the pod in execution (i.e., whose containers are running).}
    \label{fig:offloading}
\end{figure}

The overall offloading process can be summarized as follows (cf. \cref{fig:offloading}).
First, a user requests the execution of a new pod, either directly or through higher level abstractions (such as \emph{Deployments}), which is then assigned by the Kubernetes scheduler to a virtual node.
The corresponding VK instance takes charge of it, creating its twin copy in the remote cluster.
However, simply deploying pods remotely would possibly lead to resiliency problems in case of split-brain scenarios (e.g., due to temporary connectivity loss between clusters), causing service disruption if the remote pods were deleted following node failure or eviction.
For this reason, \emph{liqo} resorts to the remote creation of a \emph{ShadowPod}, a CR wrapping the pod definition and triggering the remote enforcement logic.
Ultimately, it leads to the generation of the corresponding twin pod, while transparently ensuring execution resiliency independently of the connectivity with the originating cluster.
In a nutshell, local pod operations (i.e., creations, updates and deletions) are translated to corresponding ones on remote ShadowPods, while automatic remapping is performed by the incoming reflection logic to locally propagate pod status updates in the main cluster when appropriate.

\subsubsection{Resource and service reflection}
The \emph{liqo} VK deals also with the remote propagation and synchronization of the artifacts required for proper execution of the offloaded workloads.
The reflection process is enabled by system administrators on a per-\emph{namespace} basis, together with pod offloading (cf. \cref{sec:liqo:policies}). Yet, specific artifacts (e.g., sensitive secrets) can be manually annotated and excluded.
Currently, it supports shadow \emph{ConfigMaps} and \emph{Secrets}, which typically hold application configs, as well as shadow \emph{Services} and \emph{EndpointSlices (epslices)}, to allow for intercommunication between microservices spread across multiple clusters.

\begin{figure}
    \centering
    \includegraphics{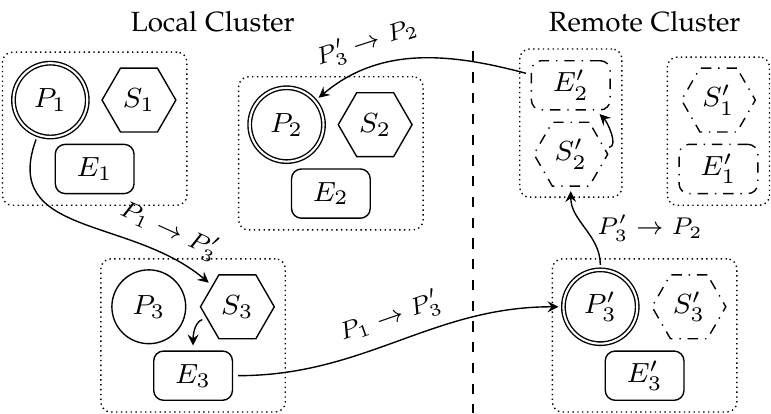}
    \caption{Graphical representation of the communication patterns between three microservices spread across two different clusters through \emph{liqo}. Dashed polygons represent shadow resources, while double circles indicate that the pod is actually in execution.}
    \label{fig:exposition}
\end{figure}

In this respect, let consider the example shown in \cref{fig:exposition}: a given application $A$ is composed of three microservices (i.e., pods), namely $P_1$, $P_2$ and $P_3$, exposed through the respective service $S_i$, in turn associated with epslice\footnote{For the sake of dissertation, we assume here a single \textit{epslice} per service, although there may be multiple (mostly for scalability reasons). Indeed, this possibility is leveraged by \emph{liqo} to segregate the reflected entries from the ones referring to local pods and achieve better scalability.} $E_i$.
Additionally, let assume $P_1$ and $P_2$ are executed on local workers, while $P_3 \equiv P^{\prime}_3$ is offloaded to a remote cluster through a virtual node.  
Once the \emph{liqo} network fabric is configured to allow inter-cluster pod-to-pod communication (cf. \cref{sec:liqo:network-fabric}), $P_1$ can directly contact $P^{\prime}_3$ through the corresponding service\footnote{Technically speaking, pods can communicate directly even without exploiting the service abstraction. Yet, the latter is typically leveraged to define a single point of access agnostic from the underlying pods and supporting DNS discovery mechanisms.} $S_3$.
As a matter of fact, the local Kubernetes control plane perceives the remote pod as executed locally and, given its (possibly remapped) IP address is present as part of its status thanks to the incoming reflection process, it creates the corresponding epslice entry (i.e., $E_3$) as usual to allow traffic forwarding.
In the opposite scenario (e.g., the remote pod $P^{\prime}_3$ willing to communicate with a local one --- $P_2$), the outgoing reflection takes action.
First, it creates the shadow copy $S^{\prime}_i$ of the local services, to enable transparent DNS discovery without requiring IP correspondence.
Second, it configures the appropriate epslice entries (possibly remapping the IP addresses, according to the network fabric configuration) to account for the local service endpoints: indeed, these cannot be managed automatically by Kubernetes, as the corresponding pods are not physically present in the remote cluster.
In the end, when $P^{\prime}_3$ contacts $S^{\prime}_2 \equiv S_2$, the standard logic forwards the request to one of the IP addresses present in the epslice $E^{\prime}_2$, eventually reaching the local pod through the \emph{liqo} network fabric, which performs the appropriate NAT translations if necessary.
Multiple replicas of the same microservice spread across different clusters, and backed by the same service $S_x$, are also handled transparently.
Indeed, each pod, no matter where it is located, contributes with a distinct epslice entry, either by the standard control plane or through outgoing reflection, hence becoming eligible during the service load-balancing process (possibly leveraging standard Kubernetes mechanisms to favor traffic locality and reduce inter-cluster communication).

\subsection{Workload Scheduling Policies}\label{sec:liqo:policies}

Each peered remote cluster is associated with a set of labels, key/value pairs describing its main characteristics (e.g., the geographical region, the hosting provider, etc.), configured by its administrators and automatically propagated to the corresponding virtual node.
This allows for fine-tuned selection of the cluster(s) each workload shall be executed on, according to its requirements and the resource continuum capabilities.
Specifically, \emph{liqo} provides a two-levels selection mechanism.
First, administrators can enable remote offloading, along with resource reflection, on a per-\emph{namespace} basis, while possibly selecting for each one a specific subset of remote clusters through their distinguishing labels (e.g., requiring those in a certain country).
Advanced configurations are foreseen to tune namespace remapping for collisions handling, as well as possibly preventing workloads scheduling on local nodes or, vice versa, preferring local to remote nodes unless in case of excessive cluster load.
Second, additional constraints can be configured at deploy time, to further restrict the eligible targets for each workload based on their requirements (e.g., enforcing front-end components to be hosted close to the end users, while introducing no additional constraints for back-end workloads); hence, fostering the intent-driven approach required by liquid computing.
Under the hood, each requirement is mapped to standard Kubernetes mechanisms (i.e., \emph{taints/tolerations} and \emph{affinities}), to comply with established practice and traditional operational procedures. 

\subsection{The Liqo Network Fabric}\label{sec:liqo:network-fabric}

The \emph{liqo} network fabric is in charge of transparently extending the Kubernetes network model across multiple independent clusters, such that offloaded pods can communicate with each other as if they were all executed locally. 
Traditionally, Kubernetes guarantees that pods on a node can communicate with all pods on any node without NAT translation.
\emph{liqo} broadens this requirement, ensuring all pods in a given cluster can communicate with all pods on all remote peered clusters, either with or without NAT translation.
Indeed, the transparent support for arbitrary clusters, with completely uncoordinated parameters and components (e.g., CNI) makes impossible to guarantee non-overlapping pod IP address ranges (i.e., \texttt{PodCIDR}).
This requires the support for IP translation mechanisms, provided that NAT-less communication is preferred whenever address ranges are disjointed.
Following industry-standard practice, the clusters interconnection is delegated to secure VPN tunnels, which are dynamically established at the end of the peering process.

\begin{figure}
    \centering
    \includegraphics{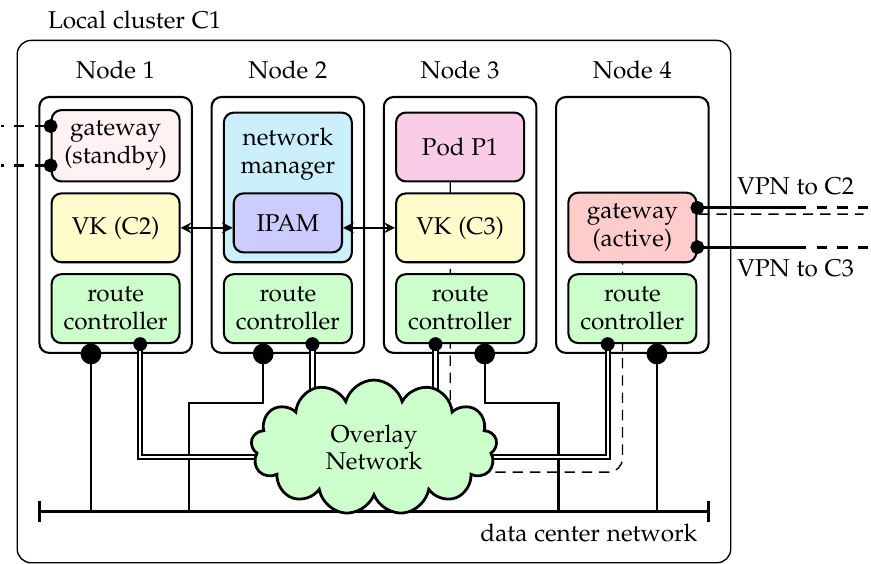}
    \caption{Main \emph{liqo} components of the network fabric subsystem, including the interconnection between clusters. The dashed line shows the path followed by traffic originating from a pod P1 and directed to one hosted on cluster C2, flowing through the overlay network to reach the gateway pod and eventually entering the VPN tunnel.}
    \label{fig:network-fabric}
\end{figure}

The \emph{liqo} network fabric currently implements a \emph{hub and spoke} topology\footnote{Ongoing work is focusing on the implementation of the opportunistic mesh configuration, to provide direct connectivity between peripheral clusters when appropriate.} (cf. \cref{sec:pillars:network-continuum}) and it is composed of three main components (\cref{fig:network-fabric}).
First, the \emph{network manager}, responsible for negotiating the connection parameters (i.e., VPN technology, IP address ranges, etc.) with each remote cluster through the exchange of appropriate CRs.
It features also an IP Address Management (IPAM) plugin, which deals with possible network conflicts through the definition of high-level NAT rules (enforced by the \emph{gateway}), while also exposing an interface consumed by the \emph{liqo} VK reflection logic to handle IP addresses remapping.
This can occur in case of overlapping \texttt{PodCIDRs} between clusters, which are transparently managed through a 1:1 translation rule to a second equivalent address range negotiated at peering time.
Additionally, ad-hoc remapping to an external free pool of IP addresses is also foreseen to support the communication between two arbitrary pods (through the respective services, hence \emph{epslices}) hosted by two different peripheral clusters in case of indirect address conflicts.

The second component is the \emph{gateway}, in charge of the setup of the VPN tunnels towards remote clusters, based on the negotiated parameters.
It implements a generic southbound interface to allow for multiple underlying drivers, although \emph{liqo} currently supports only the \emph{WireGuard}~\cite{DONENFELD2017} plugin, a modern VPN solution with state-of-the-art cryptography.
Additionally, it appropriately populates the routing table, as well as translates and installs, leveraging \emph{iptables}, the different NAT rules requested by the network manager. 
Although this component is executed in the \emph{host network}, as dealing with the node networking stack, it relies on a separate network namespace and policy routing to ensure isolation and prevent conflicts with the Kubernetes CNI plugin.
The gateway supports \emph{active/standby} high-availability, to ensure minimum downtime in case the main replica is restarted.

Finally, the third element is the \emph{route controller}, a \emph{DaemonSet} (i.e., a component executed on all physical nodes of the cluster) that is responsible for configuring the appropriate routing entries, and possibly an overlay network, to forward all traffic from local pods/nodes and directed to remote clusters through the gateway (and thus the VPN tunnel).
Once more, the high-level control logic leverages an abstract southbound interface, to allow for multiple underlying technologies.
Specifically, \emph{liqo} currently supports both a direct routing setup, hence leveraging the native infrastructure whenever possible, and a \emph{VXLAN}-based overlay network, for scenarios incompatible with the previous approach.

\subsection{The Liqo Storage Fabric}\label{sec:liqo:storage}

Along with the support of stateless workloads, \emph{liqo} transparently enables the offloading of stateful tasks through a transparent inter-cluster storage continuum.
This feature is enabled by the storage fabric subsystem, which tackles the problem through two different techniques.
First, adopting the \emph{data gravity approach} detailed in \cref{sec:pillars:storage-continuum}: whenever a workload is required to access an already existing pool of storage, a set of automatic policies forces its execution in the appropriate cluster.
Second, deferring storage binding until its first consumer is assigned to a given cluster, thus ensuring new storage pools are created in the exact location where their associated workloads have been scheduled to.

Albeit simple, these approaches extend standard Kubernetes practice to the entire resource continuum, as well as they fulfill most common use-cases.
Let consider first a high availability and disaster recovery scenario, with a database instance that needs to be replicated among different clusters.
Multiple member replicas can be spawned leveraging traditional in-cluster mechanisms (typically relying on the \emph{StatefulSet} abstraction), while configuring at the same time the appropriate intent-driven policies to enforce spreading across different virtual nodes (i.e., clusters).
Upon scheduling, a new storage pool is created in the appropriate location, and associated with each replica, which will continue to be attracted by that virtual node even following subsequent restarts.
As a second representative example, let consider an existing storage pool, attached to either a local or a remote cluster, which contains the data to be processed by a batch job.
Upon creation, the job is automatically constrained to be executed by the (virtual) node owning the corresponding piece of storage.
Hence, ensuring it can be accessed directly, without the need for expensive copy operations and enforcing at the same time data locality, which might be required by law regulations or corporate policies.

\begin{figure}
    \centering
	\subfloat[]{\includegraphics{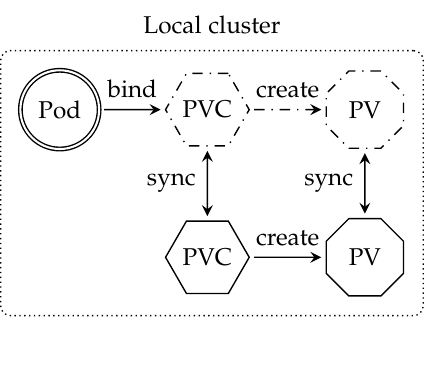}\label{fig:storage-fabric-local}}
	\hfill
	\subfloat[]{\includegraphics{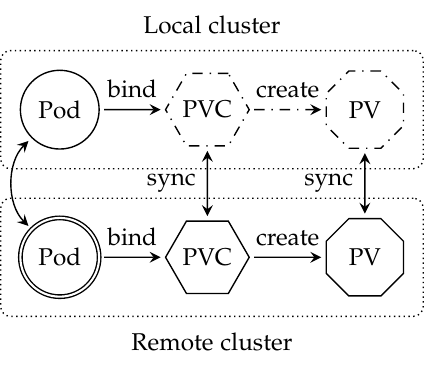}\label{fig:storage-fabric-remote}}
	\caption{Graphical representation of the persistent storage provisioning logic, in case the binding pod is scheduled on
    \protect\subref{fig:storage-fabric-local} a physical node or
	\protect\subref{fig:storage-fabric-remote} a virtual node.
	Dashed polygons represent virtual resources, while double circles indicate that the pod is actually in execution.}
	\label{fig:storage-fabric}
\end{figure}

Kubernetes leverages the \emph{PersistentVolumeClaim} (PVC) abstraction to represent a request for storage by a user, which eventually leads to the provisioning of a \emph{PersistentVolume} (PV) (i.e., the actual piece of storage a pod can bind to), either manually or through a \emph{StorageClass}.
In this context, \emph{liqo} implements a \emph{virtual} storage class, which embeds the logic to create the appropriate storage pools on the different clusters.
Whenever a new PVC associated with the virtual storage class is created, and its consumer is bound to a (possibly virtual) node, the \emph{liqo} logic goes into action (cf. \cref{fig:storage-fabric}).
If the target is a physical node, PVC operations are then remapped to a second one, associated with the corresponding \emph{real} storage class, to transparently provision the requested volume.
Differently, in case of virtual nodes, the reflection logic is responsible for creating the remote shadow PVC, remapped to the negotiated storage class, and synchronizing the PV information, to allow pod binding.
Finally, locality constraints are automatically embedded within the reflected PVs, to force each workload to be scheduled only on the clusters where the associated storage pools are available.

\section{Implementation Details}\label{sec:implementation}

We have implemented \emph{liqo} in about \num{30000} lines of Go.
According to standard practice in Kubernetes, we leveraged Custom Resource Definitions (CRDs) to describe the user-facing APIs for \emph{liqo} configuration and its internal status. 
Overall, we defined more than ten new APIs, describing the remote discovered clusters, along with the desired peering status, dealing with resource and network parameters negotiation, as well as expressing namespace offloading policies and supporting resilient remote pod execution.  
The business logic is implemented in accordance with the standard Kubernetes \emph{operators} pattern, with each controller responsible for enforcing the observed status in the cluster to match the desired one expressed by means of the corresponding resource (i.e., CR).
This paradigm guarantees separation of concerns between each component (i.e., each controller deals with a single resource, and it is responsible for a precise subset of operations), while the control loop-driven approach ensures that failures are automatically corrected, eventually reaching the desired state.
Feedback is returned to the administrators according to standard approaches, updating the \emph{status} stanza of the corresponding resource and through Kubernetes events.
Each operation performed by the controllers (e.g., creation or update of existing resources) is idempotent, ensuring that temporary errors and component restarts are handled correctly, without undesired side effects.

The overall \emph{liqo} code-base is subdivided in multiple cooperating components, each one packaged as a separate Docker container and executed by the hosting Kubernetes cluster. 
Besides the network fabric detailed in \cref{sec:liqo:network-fabric}, \emph{liqo} includes the following four components.

\emph{liqo-controller-manager:} it groups together the main operators dealing with \emph{liqo} resources.
We leveraged the controller runtime project~\cite{CTRLRUNTIME}, an abstraction built on top of the Kubernetes client to streamline the implementation of operators and efficiently use shared object caches to reduce the interactions with the API server.

\emph{liqo-virtual-kubelet:} executed in one replica for each remote cluster, ensuring isolation and segregating the different authentication tokens, it is responsible for the lifecycle of the virtual node and of the pods therein hosted, as well as for resource and service reflection (cf. \cref{sec:liqo:virtual-node}). 
Being the key component responsible for the computation offloading performance, it is implemented leveraging lower-level concepts such as \emph{informers} and \emph{working queues}~\cite{HAUSENBLAS2019} (i.e., the operators building blocks, as also done by core Kubernetes components) to increase the control and reduce possible penalties inside the offloading \emph{fast path}, regardless of the number of objects processed in parallel.
Each resource type (e.g., pods, services, epslices, ...) is associated with a custom reflection routine, accounting for parameters remapping and reduced information sharing.
Finally, smart caching mechanisms limit the interactions with the Kubernetes API server and with other \emph{liqo} components (e.g., to handle pods and epslices address translation).

\emph{liqo-crd-replicator:} component responsible for the interaction between peering clusters, enabling resource negotiation and network setup procedures through the exchange of CRs. 
It leverages a custom manager module, starting and stopping the resource synchronization logic towards each remote cluster based on the current peering phase and the overall configuration. 
Resource synchronization is implemented through a generic routine that enters in action whenever either the local or the remote version of a marked resource is modified (as detected by Kubernetes \emph{informers}), and realigns the two digital twins, with the local copy being the source of truth for the \emph{spec} stanza, and the remote one for the \emph{status} stanza. 
Hence, it transparently implements back and forth communication protocols through Kubernetes CRs.

\emph{liqo-webhook:} a mutating webhook enabling the appropriate subset of pods to be potentially scheduled on virtual nodes, based on the configured high-level policies.
Specifically, this is performed enriching the pods specification with the appropriate \emph{toleration} for the virtual nodes \emph{taint}, as well as introducing additional node \emph{affinity} constraints.

The deployment of the \emph{liqo} components is managed through a Helm chart, as per standard practice. 
Furthermore, \emph{liqo} features also a CLI tool (\emph{liqoctl}) that streamlines its configuration, automatically retrieving the appropriate parameters depending on the underlying environment (e.g., cloud provider, network setup). 
Additionally, it simplifies the manual definition of peering candidates and the selection of local namespaces for offloading to remote clusters, along with the specification of possibly complex policies.
Finally, \emph{liqo} is compatible with on-premise Kubernetes clusters (both vanilla and OpenShift-based), managed clusters hosted on major cloud providers, including Amazon EKS, Microsoft AKS and Google GCP platforms, and lightweight distributions, such as \emph{k3s}. 
Further details about the full compatibility matrix are available in the official online documentation.
\section{Experimental Evaluation}\label{sec:validation}

This section presents the experimental evaluation of the most recent version of \emph{liqo} at the time of writing (v0.4.0) as an insight of the potential performance and scalability properties of the liquid computing paradigm, taking into account the more limited scope of the current implementation.

\subsection{Peering Establishment}\label{sec:validation:peering}

Given the \emph{fluid topology} (\cref{sec:vision:characteristics}) characteristic of liquid computing, with a potential huge number of (short-living) peers, this test assesses the scalability of the peering establishment process, to evaluate the time elapsing from the discovery of a new peering candidate to the creation of the associated virtual node, while varying the number of target clusters.
The testbed consists of $n$ Kubernetes clusters in the \emph{super cluster} configuration, with a central entity (\emph{hub} cluster) establishing uni-directional peerings towards all peripheral clusters.
To simplify the setup and tear down of the entire testbed, as well as guaranteeing the replicability of the experiments, each cluster is implemented by a \emph{k3s} (v1.21.3-k3s1) instance executed within a Docker container, all together hosted by a single Kubernetes cluster.\footnote{The hosting Kubernetes cluster was composed of ten worker nodes, each characterized by 16 virtual cores and \SI{64}{\giga\Byte} of RAM.} 
Each \emph{k3s} cluster is by no means characterized by reduced functionality compared to a bare-server installation, while \emph{k3s} itself likely represents a privileged distribution for edge-oriented scenarios, thanks to its reduced demands in terms of computing resources.
To further reduce possible interference between the different instances, we leveraged the more performance-oriented \emph{etcd} database (instead of the \emph{k3s} default, \emph{SQLite}) and mounted its directory to a RAM-backed file-system, hence, preventing concurrent disk access bottlenecks when increasing the number of clusters hosted by the same physical worker. 
Similarly, relevant Docker images are retrieved in advance, to prevent their download from the Internet during the actual tests. 
All measurements have been performed through a custom tool executed on the hub cluster, which is responsible for identifying the peripheral clusters and starting the peering process (i.e., creating the corresponding \emph{ForeignCluster} resource), while monitoring the time required to complete each of the different peering phases.
The complete artifacts required to replicate the setup and perform the measurements are available on GitHub.\footnote{\url{https://github.com/liqotech/liqo-benchmarks/}}

\begin{figure}
    \centering
    \includegraphics{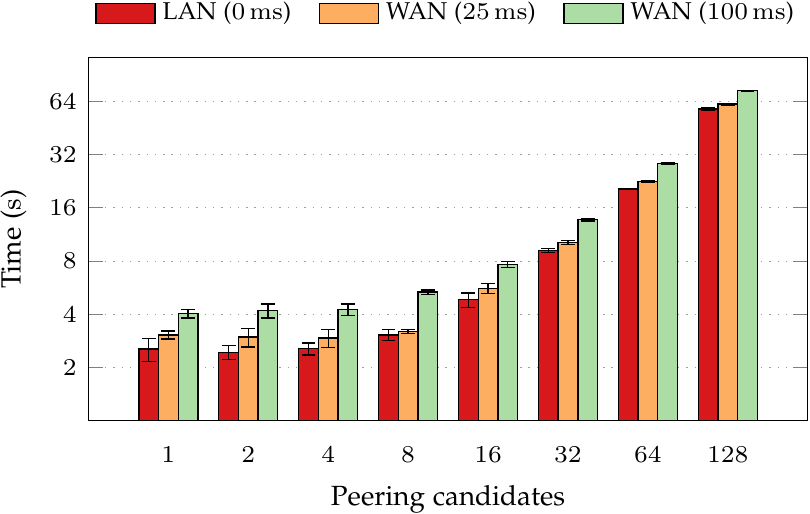}
    \caption{Peering establishment performance varying the number of peripheral clusters, and the latency with respect to the hub.}
    \label{fig:bench-peering}
\end{figure}

\Cref{fig:bench-peering} presents the outcome of the benchmark, displaying the time elapsed from the beginning of the peering process up to its completion (i.e., when all the virtual nodes are ready for application offloading), for a number of peripheral clusters ranging from 1 to 128.
To assess the impact of the distance between the hub and the peering candidates, we consider three different scenarios:\footnote{Additional latency is emulated on the hub cluster through \emph{netem}.}
\begin{enumerate*}[label=(\textit{\roman*)}]
    \item negligible latency, with all clusters located in the same LAN;
    \item \SI{25}{\milli\second} RTT latency, compatible with different sites spread across Europe;
    \item \SI{100}{\milli\second} RTT latency, accounting for intercontinental links.
\end{enumerate*}
All measurements are repeated ten times, with the error bars representing the resulting standard deviation. 
Results show that the total time increases mostly linearly with the number of parallel peering candidates, while being characterized by a constant lower bound when dealing with less than ten clusters.
The overall trend is consistent regardless of the underlying network latency, with the European scenario introducing a \SIrange[range-phrase=\,--\,]{10}{20}{\percent} overhead and the intercontinental one being associated with a relatively higher burden (in both cases especially when performing few peering establishments in parallel).

\begin{figure}
    \centering
    \includegraphics{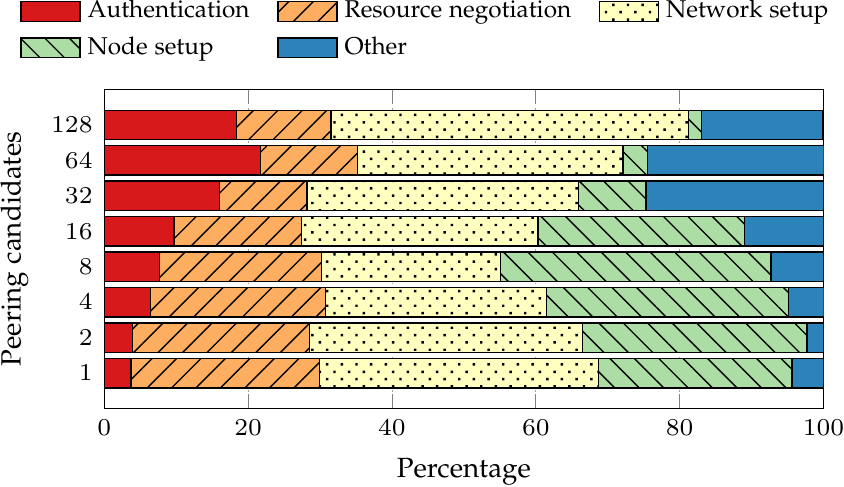}
    \caption{Peering establishment performance varying the number of peripheral clusters, break down by component (on-LAN scenario).}
    \label{fig:bench-peering-breakdown}
\end{figure}

The breakdown of the previous numbers according to the most important steps and averaged across all peering sessions performed in parallel (subject to the degree of parallelism enabled by the different \emph{liqo} components), is presented in \cref{fig:bench-peering-breakdown}; results remain mostly constant during the entire evaluation, with authentication, resource negotiation and network setup being the most demanding steps in all the considered scenarios.
Indeed, the first requires computationally expensive cryptographic operations, while the others involve parameter negotiations between the two peering clusters, which become even more prominent in case of the intercontinental scenario due to the increased network latency. 
The node setup, which is started in parallel to the network setup, impacts primarily in case of few peering candidates, with its total time in case of large number of peerings being marginally larger than the network setup.
In this case, on the other hand, the \emph{other} phase, which represents the time required for information propagation downstream the peering pipeline, gains relevance because the subsequent steps are busy processing different candidates.
Overall, these results confirm the scalability of the \emph{liqo} peering process, which required way less than one second for each target cluster in the most demanding scenario.
Finally, overall numbers may be further reduced by tuning the parallelism of the \emph{liqo} logic, although at the expense of an increased resource consumption which, currently represents a very limited cost (more details in \cref{sec:validation:resource}).

\subsection{Application Offloading}
The second benchmark analyzes the capability to start a huge burst of pods, which may be impacted by the hierarchical scheduling capabilities of \emph{liqo}.
We compared the pod startup time in vanilla Kubernetes, \emph{liqo}, and alternative open-source solutions such as \emph{Admiralty} (v0.14.1) and \emph{tensile-kube} (v0.1.1-24-g2bd91c2). Both projects also leverage the VK abstraction, although adopting different approaches under the hood (cf. \cref{sec:related-work}).
Indeed, \emph{Admiralty} focuses on a custom scheduling logic, while \emph{tensile-kube} adopts an offloading approach similar to \emph{liqo}, but it implements no remote resiliency mechanisms to prevent split brain scenarios.
It is worth mentioning that neither solution includes an automatic peering mechanism (as far as their open-source version is concerned), and thus have not been considered in the benchmark in \cref{sec:validation:peering}.

We leveraged a testbed composed of two \emph{k3s} clusters (executed within a container as in the previous case), one playing the role of the \emph{resource provider}, and the other of the \emph{consumer}, hence sticking to the \emph{elastic cluster} scenario.\footnote{The testbed was hosted by a Kubernetes cluster composed of six worker nodes, totally encompassing 332 virtual cores and \SI{2}{\tera\Byte} of RAM.}
For scalability reasons, worker nodes (i.e., those actually executing the offloaded applications) are represented by \emph{kubemark hollow nodes}~\cite{KUBEMARK}, which are backed by a component, named \emph{hollow kubelet}, executed in its own container, and that pretends to be an ordinary kubelet, but it does not start any container it is assigned to, it just lies it does.
This allows to start a massive number of (fake) containers, even tens of thousands (i.e., comparable with the maximum number of pods supported by Kubernetes~\cite{KUBERNETESLARGE}, $\approx \SI{150}{\kilo\nothing}$), with limited resource demands given that containers are not actually running.
This does not invalidate the results, given the reduced startup time affects all solutions equally; vice versa, it better highlights the possible overheads introduced by the offloading process, compared to vanilla Kubernetes.
All hollow kubelets (100 in our setup) connect to the provider, registering as an additional node.
We adopted the official hollow kubelet code base available upstream (v1.21.4), with a simple modification to assign pods an IP from the correct \texttt{PodCIDR}, instead of a fake one, for increased realism.
All measurements have been performed through a custom tool, which creates the appropriate deployments and waits for the corresponding pods to be generated, possibly offloaded, and become ready.
We initially executed it directly on the resource provider cluster to determine the vanilla Kubernetes baseline, and then on the consumer (each time peered through a different technology with the provider) to assess the offloading performance.

\begin{figure}
    \centering
    \includegraphics{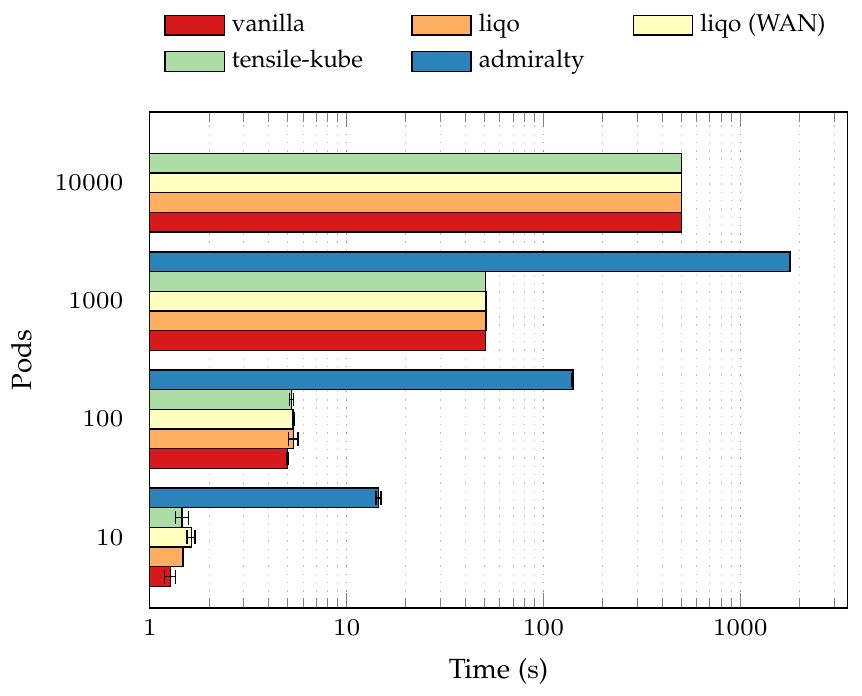}
    \caption{Application offloading performance comparison, varying the number of pods to be started. The \emph{Admiralty} result is not included for the \SI{10}{\kilo\nothing} pods case, given it is largely out of scale.}
    \label{fig:bench-offloading}
\end{figure}

\Cref{fig:bench-offloading} presents the outcome of the evaluation, depicting the time elapsed from the creation of a deployment up to the instant all generated pods are effectively ready, for a number of pods varying between \num{10} and \num{10000}.
All measurements are repeated ten times, and the error bars represent the resulting standard deviation.
The outcome of the benchmark is twofold. 
On the one hand, both \emph{liqo} and \emph{tensile-kube} displayed excellent performance, introducing practically no overhead compared to vanilla Kubernetes.
Still, \emph{liqo} supports additional mechanisms to ensure application reliability even in case of split brain scenario, which \emph{tensile-kube} does not.
Differently, the scheduling-driven approach adopted by \emph{Admiralty} turned out to be associated with much worse performance, introducing unbearable overhead when offloading a high number of pods at the same time.

As for \emph{liqo}, we additionally evaluated the performance in case consumer and provider are interconnected by a high-latency (\SI{100}{\milli\second}) WAN: no relevant difference emerges compared to the on-LAN scenario, with a slight overhead visible only when offloading \num{10} pods in parallel (approx. \SI{100}{\milli\second}).
Finally, we varied the number of deployments originating the target set of pods, accounting for multiple use-cases (i.e., ranging from a massively replicated monolithic workload, to a complex application composed of a hundred microservices, each with a significant number of replicas).
No significant outcome emerged in addition to the previous considerations, with the difference between vanilla Kubernetes and \emph{liqo} being always smaller than the error bands.
Hence, we have omitted these results from \cref{fig:bench-offloading} for the sake of conciseness.

\subsection{Service Exposition}
This first test evaluates the time required by the \emph{liqo} reflection logic to replicate a service and all the associated \emph{epslices} to a remote cluster, hence making them available for consumption by remote applications, but without including the time needed for the vanilla Kubernetes data plane configuration (e.g., kube-proxy).
This highlights the time required to propagate a new service (or a new running endpoint) across the control plane of the virtual cluster as well as the scalability of the solution.
We leveraged a testbed similar to the previous one, characterized by two \emph{k3s} clusters and a set of hollow nodes to host the fake containers.
However, we considered the symmetric scenario, with a varying number of pods started locally and, once ready, exposed through a single Kubernetes service.
A custom tool is responsible for measuring the time required to fill all epslice entries, both on the local cluster (i.e., by the vanilla Kubernetes logic) and on the remote one (i.e., through the \emph{liqo} reflection logic).

\begin{figure}
    \centering
    \includegraphics{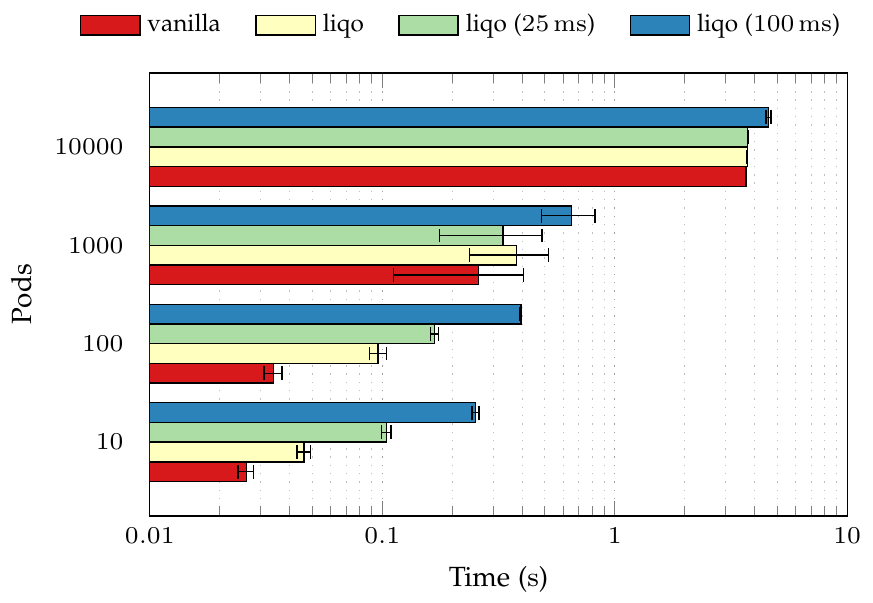}
    \caption{Service exposition performance comparison, varying the number of endpoint pods and the inter-cluster latency.}
    \label{fig:bench-exposition}
\end{figure}

The outcome of the benchmark is depicted in \cref{fig:bench-exposition}, which shows the ten-runs average of the time elapsed from the creation of a service targeting the given number of pods to the complete creation of the corresponding epslices (marked as \emph{Ready}).
The graph confirms the limited performance overhead introduced by the \emph{liqo} reflection logic compared to vanilla Kubernetes, accounting for a few milliseconds only even in the most demanding scenario.
Given the overall short times required to complete the process, the effect of the underlying network latency, both considering the European (\SI{25}{\milli\second}) and the intercontinental (\SI{100}{\milli\second}) scenarios, becomes relevant.
Yet, in absolute terms, the overhead is definitely close to the network latency itself, which is unavoidable.

To further characterize the service propagation overhead, we additionally measured the time elapsed from the creation of a service to the instant it is fully reachable; 
therefore, including the reflection logic, the configuration of \emph{iptables} rules by vanilla kube-proxy, and the network fabric data plane contribution (e.g., packets traversing the \emph{WireGuard} tunnel).
In this scenario, we leveraged a single \emph{nginx} pod as service endpoint, with a custom tool executed in both clusters and continuously probing the service through TCP SYN segments, until the corresponding acknowledgement is received; hence, confirming the service is reachable. 
Across ten runs, the local service (i.e., where the back-end pod is running) became accessible in \SI[separate-uncertainty=true, multi-part-units=single]{0.091(12)}{\second}.
As for the remote cluster, the \emph{liqo}-reflected service turned reachable in \SI[separate-uncertainty=true, multi-part-units=single]{0.100(8)}{\second} (in case of negligible inter-cluster latency), and \SI[separate-uncertainty=true, multi-part-units=single]{0.218(32)}{\second} in the WAN (\SI{100}{\milli\second}) scenario.
Overall, showing once more limited overhead compared to vanilla Kubernetes, despite the increased functionality.

Focusing finally on network throughput, the data plane handling the actual communication between any two pods hosted by different clusters relies on standard VPN technologies (i.e., WireGuard) and inherits their performance, as well as those of the underlying network.

\subsection{Stateful workloads}

This benchmark assesses the performance of the \emph{liqo} storage fabric subsystem, concerning both the creation of new PVs and the binding of a pod to an already existing volume.
We adopted a testbed composed of two \emph{k3s} clusters, complemented by a custom tool responsible for the creation of a \emph{StatefulSet} (i.e., the Kubernetes abstraction representing a set of pods with consistent identities, each characterized by one or more volume claims) and the measurement of the pod startup time (including volume creation and binding).
Concerning persistent volumes management, we evaluated the usage of a vanilla storage class (i.e., the one included by default with \emph{k3s}), as well as of the \emph{liqo}-provided one, when pods are hosted by either the local or the remote cluster.

\begin{figure}
    \centering
    \includegraphics{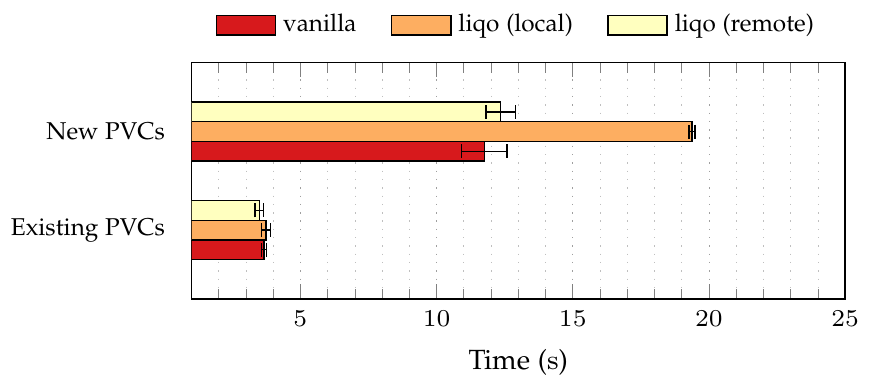}
    \caption{Stateful workloads startup performance comparison, both when the underlying storage needs to be created and when it already exists.}
    \label{fig:bench-stateful}
\end{figure}

\Cref{fig:bench-stateful} presents the ten-runs average of the startup time in case of a \emph{StatefulSet} originating five replicas,\footnote{The \emph{StatefulSet} was configured to start all replicas in parallel, rather than sequentially, to better stress the storage subsystem.} hence mimicking a high-availability database setup.
First, we analyzed the initial deployment of the application (\emph{New PVCs} in figure), which includes the creation of the PVCs, that of the underlying PVs, and the startup of the pods themselves.
Results associated with the vanilla and \emph{liqo} (remote) scenarios are aligned, while the setup of volumes on the local cluster through \emph{liqo} turned out to be slower.
This limitation traces back to an external library we leveraged, which adopts by default a rather long polling period to detect the creation of the actual PVs. 
While better performance could be obtained fine-tuning that component, it is worth mentioning that the creation of PVCs in cloud-provider environments is typically much slower (on the order of minutes), as well as this operation is expected to be quite infrequent.
Differently, no relevant difference emerged when binding the pods to already existing volumes (i.e., \emph{Existing PVCs}), which instead happens whenever one or more replicas are restarted. 

\subsection{Liqo Resources Characterization} \label{sec:validation:resource}

The last test characterizes the \emph{liqo} resource demands, in terms of CPU and RAM required for the control plane execution, as well as the network traffic generated by \emph{liqo} towards the remote Kubernetes API servers during the different operational phases (e.g., peering, resource offloading, etc.).
Overall, the testbed is similar to the one adopted in \cref{sec:validation:peering}, and composed of eleven \emph{k3s} clusters, one playing the role of the hub and ten behaving as peering candidates.
CPU and RAM consumption is retrieved every second on each cluster through the APIs exposed by the \emph{containerd} container runtime, while network traffic is measured on the hub by means of a custom \emph{libpcap}-based program. 

\begin{figure}
    \centering
    \includegraphics{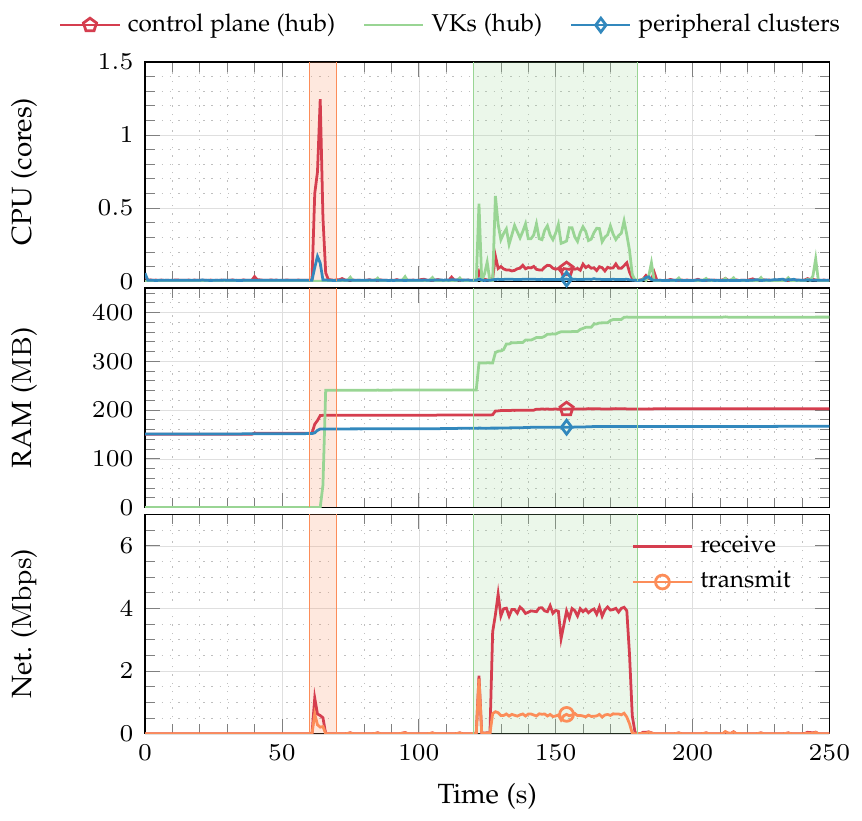}
    \caption{\emph{liqo} resource demands at rest, peering with 10 peripheral clusters (orange shaded area --- \SIrange[range-phrase=\,--\,]{60}{70}{\second}) and offloading \SI{1}{\kilo\nothing} pods (green shaded area --- \SIrange[range-phrase=\,--\,]{120}{180}{\second}). Network metrics refer to the traffic towards remote API servers, measured on the hub cluster (i.e., hosting the VKs).}
    \label{fig:bench-consumption}
\end{figure}

\Cref{fig:bench-consumption} presents the outcome of the measurements, subdivided into the \emph{liqo} control plane of the hub cluster (i.e., all the \emph{liqo} components excluding the VKs), the sum of the ten VKs hosted by the hub cluster, and the \emph{liqo} control plane of the peripheral clusters (no VKs are present in this case, since peerings are unidirectional).
As for the latter, CPU and RAM metrics are averaged (differences are not significant), hence showing the average requirements for a single cluster. 
Overall, we consider five different usage phases (highlighted with different colors in \cref{fig:bench-consumption}):
\begin{enumerate*}[label=(\textit{\roman*)}]
\item At rest and with no active peering (\SIrange[range-phrase=\,--\,]{0}{60}{\second}): in this context, \emph{liqo} requires approx. \SI{150}{\mega\Byte} of RAM on each cluster, with almost no CPU usage and zero network traffic.
\item While peering with ten peripheral clusters in parallel (\SIrange[range-phrase=\,--\,]{60}{70}{\second}) to assess the processing cost of such operation: the local control plane is characterized by a short CPU spike during the process and a few \si{\mega\Byte} increase in memory consumption, while the ten VKs, started in parallel in the hub cluster, account for approximately \SI{250}{\mega\Byte} of additional RAM in total.
The exchanged network traffic is negligible.
\item At rest and with the virtual nodes ready (\SIrange[range-phrase=\,--\,]{70}{120}{\second}): no CPU and network resources are required by \emph{liqo} to maintain the peerings active, while the memory occupancy remains stable compared to the previous phase. 
\item Offloading \SI{1}{\kilo\nothing} pods to simulate high churn rates (e.g., a large number of pods is started or changes its state) (\SIrange[range-phrase=\,--\,]{120}{180}{\second}): as for CPU usage, the ten VKs (as a whole) required a third of a CPU core, while both local and remote control plane demands remained definitely low.
VKs RAM usage increased as well, since memory consumption is directly related with the number of pods, according to the standard Kubernetes operators implementation (i.e., watched resources are cached by informers).
The information synchronization between the different clusters resulted at the same time in an additional network traffic, although it never exceeded \SI{4}{\mega\bps} in total at least in the significant experiment depicted in \cref{fig:bench-consumption}.
\item At rest, with the active peerings and the offloaded pods running on the remote clusters (\SIrange[range-phrase=\,--\,]{180}{250}{\second}): none of the considered metrics displayed variability, confirming the negligible demands in absence of transient periods.
\end{enumerate*}

Concerning the offloading phase, we repeated the entire evaluation for different numbers of pods, while keeping constant the other parameters. 
Overall, we observed similar CPU usage and network traffic, although for time frames proportional to the number of pods (e.g., $\approx \SI{30}{\second}$ with \num{500} pods), according to Kubernetes deployment pacing.
As for the VKs RAM usage, the theoretical linear correlation was not completely reflected in the actual measurements, due to the Go garbage collector behavior.\footnote{We leveraged the default GC settings during all tests.}
In our tests, each offloaded pod accounted for \SIrange[range-phrase=\,--\,]{150}{300}{\kilo\Byte} of additional RAM, with the upper range associated with lower numbers of overall pods.
In addition, different experiments done with different infrastructures (e.g., node characteristics) achieved very similar patterns, with approximately the same amount of total CPU consumed and traffic exchanged, although constrained in shorter (with more powerful nodes) or longer (with less powerful nodes) intervals. Furthermore, the above values proved to be slightly lower compared to the ones generated by ten standard kubelets controlling vanilla Kubernetes worker nodes in the same scenario.

Finally, perhaps the only metric which might deserve attention in the context of constrained devices is memory usage ($\approx \SI{160}{\mega\Byte}$ on each peripheral clusters).
Yet, it is worth mentioning that even lightweight Kubernetes distributions do require non-negligible amounts of RAM (e.g., \emph{k3s} recommends at least \SI{1}{\giga\Byte}~\cite{K3SREQUIREMENTS}, which is also confirmed by the measurements in \cite{BOHM2021}), as well as, at the time of writing, \emph{liqo} could be further optimized to reduce its demands. 

\subsection{Additional Considerations}
Our experimental evaluation demonstrates the extremely limited overhead introduced by \emph{liqo} in terms of additional resource demands and with respect to vanilla Kubernetes.
Hence, justifying the sustainability of the resource continuum abstraction, with its distinguishing characteristics, and of the building blocks enabling the scenarios detailed in \cref{sec:vision:scenarios}.
In the following, we first present a production environment benefiting from the elastic cluster scenario enabled by \emph{liqo}, and then discuss its potential to overcome Kubernetes scalability limitations, as well as service reliability aspects.

\subsubsection{Real Liqo deployment: job bursting for online exams}

Politecnico di Torino recently hosted different computer science exam sessions on CrownLabs~\cite{CrownLabs}, an open-source project started during the coronavirus pandemic onset to deliver remote computing laboratories, later extended and integrated with the official exams platform of our university. 
It allows each candidate to access her own dedicated remote application instance (started on-demand and executed as a container) of the desired environment (e.g., a full-fledged IDE such as PyCharm) from a standard web browser, while providing automatic project delivery and enforcing the appropriate restrictions to prevent cheating.
However, the Kubernetes cluster typically hosting the CrownLabs user instances was not capable enough to sustain the foreseen number of students (\numrange[range-phrase=\,--\,]{500}{600} per round, multiple rounds per day) with the desired resources (i.e., 1 CPU reserved to each instance).
The elastic cluster scenario enabled by \emph{liqo} allows part of the instances to be transparently offloaded to a secondary cluster located in a different campus area, while requiring no modifications to the CrownLabs control plane.
From extensive monitoring data, no significant performance differences emerged between the creation of local and remote instances, with the \emph{liqo} network fabric handling an average cross-cluster user traffic of \SI{100}{\mega\bps} and the storage continuum ensuring data persistence even in case of instance restarts. 
Agility proved to be one of the most distinguishing features, allowing to reserve the remote resources only during the actual exam sessions, and immediately releasing them for different purposes at the end.
Conversely, standard practice would require to either allocate additional servers to physically extend the cluster, or migrate the entire service to a larger infrastructure, with both alternatives introducing high organizational and operational overheads.

\subsubsection{Scalability: Liqo vs. Kubernetes}

According to the official documentation~\cite{KUBERNETESLARGE}, Kubernetes is currently characterized by scalability upper bounds both in terms of supported nodes ($\approx\num{5000}$) and pods ($\approx\num{150000}$).
In the context of \emph{liqo}, these limitations mainly relate to the super cluster scenario, with a single entry point potentially controlling a large number of e.g., edge clusters.
Indeed, thanks to the combination of the virtual node abstraction and the remote enforcement logic, \emph{liqo} allows to transparently deal with cluster control planes spread geographically, supporting also scenarios with unstable network connectivity and high latency, which is not possible with vanilla Kubernetes, whose control plane is fully centralized.
At the same time, \emph{liqo} has the potential to overcome the Kubernetes node limitations, given that it abstracts an entire cluster with a single node (preventing the propagation of most remote status changes) and it supports hierarchical topologies characterized by multiple indirection levels.
For instance, a large company operating thousands of edge clusters (e.g., telco edge; energy smart grids; branch offices, etc.) might leverage regional clusters as intermediate aggregation points, in turn controlled from a single national data center.
Hence, preventing to exceed node limitations in any single cluster, while dealing with much higher numbers as a whole. 
Focusing on pod offloading, as discussed in \cref{sec:liqo:virtual-node}, \emph{liqo} currently favors full Kubernetes API transparency, which requires all pods to be \emph{virtually} present in the super cluster (i.e., where they are originally created through higher-level abstractions) and accurate status synchronization.
Although consuming no local resources, they are nonetheless present in etcd, partially counting towards the Kubernetes limits.
This is inherent in the VK approach: future work could focus on additional offloading solutions that are more suitable for high cardinality scenarios, trading off full API compliance with increased scalability.

\subsubsection{Application reliability}
As discussed in \cref{sec:liqo:virtual-node}, the \emph{liqo} VK leverages periodic healthiness checks to evaluate the reachability of the remote cluster, mapping the outcome to the node readiness property.
This perfectly resembles the behavior of vanilla Kubernetes nodes, allowing at the same time the pods therein hosted to obey to standard eviction policies to enforce application reliability.
In particular, two main parameters control the entire process, indirectly determining the maximum time frame between a remote cluster turning unreachable and the hosted pods being rescheduled in a different location. 
First, the node lease duration (default: \SI{40}{\second}): the VK periodically renews a \emph{lease} (i.e., updates an appropriate resource) to confirm it is operating correctly.
In case the check fails, the VK stops doing so and the lease expires after that period, causing Kubernetes to mark the node as unreachable.
Second, pods toleration for \textit{not ready} and \textit{unreachable} nodes (default: \SI{300}{\second}), that is the maximum interval the pod is allowed to remain bound to a problematic node, before being evicted and scheduled to a different one.
In other words, the entire process requires by default at most \SI{340}{\second}.
Depending on the specific scenario, different settings may be more appropriate: lowering both values, and in particular the toleration period (which might be even set to \SI{0}{\second}), allows for faster reactions, at the cost of potentially higher churn rates in case of temporary connectivity issues.
Differently, drastically higher toleration settings might be better suited in harsh environments, to explicitly account for temporary network partitioning while letting existing workloads to evolve independently, thanks to the remote enforcement logic, guaranteed by the control plane running in the remote cluster.
Still, tolerations are set per pod, allowing for fine-grained control and application specific settings, regardless of the specific target cluster.

In other words, upon cluster disconnection, the orchestration logic in the main cluster can be either configured to immediately re-spawn, in another location, all the services that are no longer available for the sake of service continuity, or to leave services where they are.
This accounts for either the case in which services should be always available to the users of the main cluster (hence, are re-spawn elsewhere), or the services are intended for local users of the disconnected cluster, which presumably are still able to reach their local infrastructure, even in case of unreachability of the main cluster.
In the latter case, the remote control plane features a dedicated enforcement logic that ensures improved service resiliency, guaranteeing that the potential failure of a local node leads to no service disruption thanks to automatic pod rescheduling, regardless of the connectivity with the main cluster.
This is different compared to a vanilla Kubernetes cluster encompassing nodes spread geographically, as connectivity loss in a given area would isolate that group of nodes from the central control plane, lacking the possibility for any evolution of their state.
\section{Related Work}\label{sec:related-work}

The effort towards a transparent resource continuum dates back to the eighties and the concept of distributed operating systems, aiming to abstract a set of independent, autonomous and communicating CPUs that appear to users as a single computer~\cite{TANENBAUM1985,TANENBAUM1993}.
At the same time, much work focused on a common substratum for applications execution. 
First, by means of high-level programming languages (e.g., Java~\cite{JAVA1995}) to achieve architecture neutrality, and later through containerization, providing a lightweight answer to packaging and distributing interoperable applications~\cite{PAHL2015}.
This paved the way for container orchestration platforms such as Kubernetes, abstracting the resources in a data center and implementing, to some extents, the distributed operating systems vision.
At the same time, the prominent emergence of cloud computing has led to efforts towards inter-cloud architectures, aiming for better QoS, reliability and cost efficiency~\cite{GROZEV2014,TOOSI2014}.
This, in turn, fostered the expansion of this approach towards end users, introducing paradigms such as edge computing~\cite{GARCIALOPEZ2015,SHI2016} and fog computing~\cite{BONOMI2012}.
Liquid computing extends these concepts towards a uniform infrastructural substratum for seamless distributed applications orchestration.
Dynamism is a key distinguishing factor, enabling different pools of resources (e.g., cloud-based, cloudlets~\cite{SATYANARAYAN2009} and edge devices), likely under the control of different administrative domains, to transparently participate to one, or even multiple, computing continuums.
Osmotic computing~\cite{VILLARI2019}, given its broad scope, overlaps to some extents our proposal although, to the best of our knowledge, lacking a concrete characterization and targeting specifically the IoT domain.

The reminder of this section overviews the related work focusing on control plane and networking-related multi-cluster approaches in Kubernetes, including research proposals and relevant open-source projects.
Indeed, we deem Kubernetes integration to be a key factor for smooth industry adoption.
Proprietary solutions such as Google Anthos, Azure Stack, AWS Outposts and VMware Tanzu are out of scope, as bound to a specific environment, preventing the full realization of the computing continuum vision spanning across any \emph{technological} and \emph{administrative} domain.

\subsection{Multi Cluster Kubernetes Control Plane}

Kubernetes Cluster Federation (KubeFed)~\cite{KUBEFED} represents the official community solution to the multi-cluster problem, and adopts a centralized approach to coordinate multiple clusters through an appropriate set of APIs.
It implements a single point of control abstraction, as the users interact with the host cluster only, and the KubeFed control plane takes care of propagating the modifications.
To overcome its limitations in terms of failure tolerance and number of supported clusters (as mostly focusing on large infrastructures), Larsson~\emph{et al.}~\cite{LARSSON2020B} presented their vision towards a decentralized Kubernetes federation control plane.
Their approach leverages a shared database of conflict-free replicated data types (CRDTs) to synchronize the global desired state, removing the single point of failure intrinsic in centralized solutions and foreseeing the support for thousands of federated edge clusters.
Differently, Faticanti~\emph{et al.}~\cite{FATICANTI2021} showcased the combination of KubeFed and their proprietary FogAtlas framework, the latter allowing to model distributed applications and automatically schedule the different components based on specified requirements.
Still, all these solutions explicitly target infrastructures under the same administrative domain and focus on the control plane only, while requiring external solutions for inter-cluster communication.  

The Karmada open-source project~\cite{KARMADA} adopted a different approach, leveraging a custom API Server to provide a single point of control abstraction, while mimicking the standard Kubernetes one.
Then, vanilla high-level resources, such as Deployments, do not incur in the usual workflow, but get processed by the custom controllers and, depending on additional policy constraints defined through CRs, are eventually dispatched to the target clusters.
Yet, this approach, which targets only the super cluster scenario, does not provide full Kubernetes compliance, preventing administrators the possibility to transparently deal with lower-level objects (i.e, pods) for accurate monitoring and troubleshooting operations.
As an alternative solution, the \emph{GitOps} paradigm~\cite{LIMONCELLI2018} can also enable an elementary multi-cluster control plane, allowing the distribution of different tasks, described through declarative configuration files and automatically enforced by CI/CD mechanisms, across the entire infrastructure. 
Still, workload placement is completely static, lacking the possibility to dynamically migrate or scale applications across clusters to face unexpected failures or load spikes. 
Once more, this solution requires full control over the entire infrastructure, and it considers each cluster as an isolated silo, with no transparent communication support.

A different category of approaches leverages the VK abstraction to transparently masquerade the remote clusters, while introducing no API disruption and potentially supporting multi-ownership through proper isolation and permission limitations.
Besides \emph{liqo}, two main open-source projects followed this approach.
First, \emph{Admiralty}~\cite{ADMIRALTY}, a solution enabling different multi-cluster topologies, both centralized and distributed.
It features a custom scheduling logic to select the best workload execution placement, complemented by resilient remote offloading through the custom \emph{PodChaperon} abstraction and resource reflection.
Second, \emph{tensile-kube}~\cite{CAI2020}, a project developed by Tencent Games to ensure high utilization in case of resources fragmented across multiple clusters.
It supports remote pod offloading (although with no split-brain resiliency mechanisms), as well as resource reflection, along with custom scheduling and de-scheduling extensions to deal with resource fragmentation.
Differently from \emph{liqo}, tensile-kube resorts to external mechanisms for inter-cluster networking, as well as it poses no multi-ownership constraints as the entire infrastructure is assumed to be controlled by the same organization.  
Finally, the VK abstraction has also been leveraged in a different context by FLEDGE~\cite{GOETHALS2020}, a Kubernetes-compatible lightweight solution allowing individual low-resource edge devices to join an existing cloud-based cluster. 
Conversely, \emph{liqo} targets multi-cluster scenarios, abstracting entire cluster slices as virtual nodes to enable seamless application scheduling, while adopting a decentralized approach to preserve the independence of each pool of resources.

\subsection{Multi Cluster Kubernetes Network Interconnection}

As for multi-cluster networking, we identified three main classes of solutions.
First, CNI-provided (e.g., \emph{Cilium ClusterMesh}~\cite{CLUSTERMESH2019}), hence featured by standard cluster connectivity components.
However, this approach demands for coordination between all federating clusters to leverage the same technologies and prevent addressing conflicts, which is unsuitable in case of dynamic, multi-ownership scenarios.
Second, CNI-agnostic solutions, mainly including \emph{Submariner} and \emph{Skupper}.
Submariner~\cite{SUBMARINER} enables cross-cluster layer-3 connectivity using encrypted VPN tunnels, while supporting service discovery and address conflict resolution mechanisms. 
Under the hood, it leverages a centralized, broker-based architecture to negotiate the interconnection configurations.
Skupper~\cite{SKUPPER}, on the other hand, operates at layer 7 and possibly interconnects only a subset of remote Kubernetes namespaces, instead of the entire clusters they belong to.
However, the above solutions address only the cross-cluster connectivity requirements, while leaving workload orchestration and observability across the entire resource continuum to either static approaches or other external tools.
Third, service mesh-provided: a service mesh is a dedicated infrastructure layer for handling service-to-service communication, which is typically implemented as lightweight network proxies (i.e., sidecars) deployed alongside the actual applications~\cite{LI2019}. 
Popular service mesh frameworks, including Istio~\cite{ISTIO} and Linkerd~\cite{LINKERD}, feature also multi-cluster support through dedicated proxies, routing traffic from the mesh of one cluster to another.
However, they do not implement a cross-cluster control plane, lacking a single point of entry to oversee the entire multi-cluster topology, and dynamically schedule the workloads in the best available location, regardless of the underlying infrastructure topology.
In addition, service mesh solutions require complex configurations, and they come at a high cost in terms of application and latency overhead, due to the introduction of sidecars, which may be unsuitable especially in case of edge devices~\cite{SARMIENTO2021}.
\section{Conclusions}\label{sec:conclusions}

In recent years, the irrefutable success of the cloud and edge computing paradigms has brought to cluster proliferation in both small and large organizations, as well as the deployment of orchestrated edge devices running lightweight Kubernetes flavors.
This trend inevitably leads to resource fragmentation, statically constraining applications execution to predetermined silos and introducing high operational complexity when fulfilling high availability requirements and policy compliance.

In this paper, we first advocated the opportunity for the introduction of liquid computing, a novel paradigm enabling a transparent continuum of computational resources and services on top of the underlying fragmented infrastructure.
It foresees a decentralized, fluid and peer-to-peer architecture to account for the dynamism typical of edge and IoT devices, multi-ownership, to support the interconnection between different administrative domains, as well as an intent-driven approach, simplifying the characterization of each workload with high-level policies to constrain their execution to the most appropriate location.
We believe liquid computing can simplify multi-cluster operations, through its intrinsic big cluster abstraction, and enable both resource sharing, to reduce allocation inefficiencies, and brokering scenarios, extending the highly successful IXP model to cloud and edge data center slices.
Second, we presented and characterized \emph{liqo}, an open-source project fostering the liquid computing vision through the creation of dynamic and seamless Kubernetes multi-cluster topologies.
Extensive experimental evaluations have shown the effectiveness of \emph{liqo}, both in terms of limited overhead with respect to vanilla Kubernetes and better performance compared to state of the art open-source solutions, while including at the same time more advanced features.

\section*{Acknowledgment}
The authors would like to thank all the people who contributed to our journey towards the liquid computing vision, in particular Aldo Lacuku, Alessandro Olivero, Mattia Lavacca, Dante Malagrin\`{o}, all the students at Politecnico di Torino who collaborated on this project, and all the people who trusted \emph{liqo} by running it on their production clusters.

This work was partly supported by European Union’s Horizon Europe research and innovation programme under grant agreement No 101070473, project FLUIDOS (Flexible, scaLable, secUre, and decentralIseD Operating

\IEEEtriggeratref{25}\IEEEtriggercmd{\balance}
\bibliographystyle{IEEEtran}
\bibliography{main}

\end{document}